\documentclass[superscriptaddress, aps,pre, twocolumn, floatfix]{revtex4-2}
\usepackage{graphicx}
\usepackage{dcolumn}
\usepackage{amsmath}
\usepackage{latexsym}
\usepackage{mathrsfs}
\usepackage[usenames,dvipsnames]{color}

\usepackage[normalem]{ulem}
\usepackage{xcolor}

\usepackage[breaklinks,colorlinks = true,linkcolor = blue,urlcolor=blue,citecolor=blue]{hyperref}

\definecolor{dgreen}{rgb}{0,0.7,0}
\newcommand{\be}{\begin{equation}} 
\newcommand{\ee}{\end{equation}} 
\newcommand{\bea}{\begin{eqnarray}} 
\newcommand{\eea}{\end{eqnarray}} 

\begin{document}
\title{Confined Riesz gas: From Newtonian Dynamics to Emergent Hydrodynamics}

\author{Indranil Mukherjee}
\email{indranil.mukherjee@icts.res.in}
\affiliation{International Centre for Theoretical Sciences, Tata Institute of Fundamental Research,
Bangalore 560089, India}

\author{Abhishek Dhar}
\email{abhishek.dhar@icts.res.in}
\affiliation{International Centre for Theoretical Sciences, Tata Institute of Fundamental Research,
Bangalore 560089, India}

\author{Manas Kulkarni}
\email{manas.kulkarni@icts.res.in}
\affiliation{International Centre for Theoretical Sciences, Tata Institute of Fundamental Research,
Bangalore 560089, India}

\begin{abstract}
We investigate the real-time dynamics of the Riesz gas (RG), a paradigmatic system of particles with power-law interactions that encapsulates several many-body interacting classical systems. While the equilibrium properties of the system have been extensively studied the connection between microscopic dynamics, macroscopic evolution and thermalization has remained largely unexplored. Combining extensive large-scale Newtonian simulations with a hydrodynamic framework formulated in terms of three coarse grained fields -- density, velocity, and temperature -- we establish a direct and quantitative correspondence between the two descriptions. The hydrodynamic theory incorporates two phenomenological transport coefficients, the bulk viscosity $\xi$ and thermal conductivity $\kappa$, which encode dissipative effects at a macroscopic level. By analyzing representative classes of initial conditions, such as dome-like and Newton-cradle-type profiles, we find excellent agreement between microscopic and hydrodynamic evolution over a broad range of timescales, starting from relatively early all the way up to times at which steady state is attained. Notably, this agreement persists across a substantial window of $\xi$ and $\kappa$, demonstrating the robustness of the emergent hydrodynamic description. Our results provide a systematic validation of hydrodynamics in the RG and establish a concrete bridge between Newtonian dynamics and continuum theories in systems with power-law interactions, starting from first principles.
\end{abstract}
\maketitle

\textit{Introduction.--} Systems with long-range interactions have long occupied a central position in both physics and mathematics, owing to their rich equilibrium and dynamical properties. Among these, the Riesz gas (RG) \cite{Riesz,Lewin_Riesz,Riesz_avg_density_SA}, a system of $N$ particles interacting via all-to-all pairwise potentials of the form ${\rm sgn}(k)|x_i-x_j|^{-k}$, with $k$ being the power-law exponent, has emerged as a particularly versatile and unifying framework. It interpolates continuously between several paradigmatic models, including the one-dimensional one component plasma ($k=-1$) ~\cite{Lenard_JMP_1963,Statistical_MPCPS_Baxter,dhar_IOP_2018,dhar_PRL_2017}, Dyson's log-gas ($k\to 0 $)~\cite{Dyson_JMP_1962,dyson_JKP_1962,mehta2004random,forrester2010log}, integrable Calogero model ($k=2$)~\cite{Calogero_JMP_1969,Calogero_JMP_1971,Calogero_LNC_1975,OLSHANETSKY1981, polychronakos_IOP_2006} and the hard rod gas ($k\to \infty$)~\cite{HR_Tonks,JK_HardRods}. There have been generalizations of the RG to finite-ranged cases, where the interaction is with $d$ nearest neighbours and not all-to-all~\cite{finite_range_Riesz_MK}. In addition to the pairwise potentials of power-law form these systems are often placed in an additional external confining trap, mostly quadratic in nature. Recent advances have also highlighted potential experimental relevance in cold atoms~\cite{James1998,brown_CUP_2003,chalony_PRA_2013,zhang_nature_2017} and related platforms, where tunable long or finite ranged interactions of power-law form are increasingly accessible. 

In recent years, there has been significant progress in understanding the collective and equilibrium properties of the RG. 
The family of RG and its variants provides a rare setting where exact results on large $N$ density profiles~\cite{Riesz_avg_density_SA}, gap statistics~\cite{Saikat_Riesz_2022}, large deviation principles~\cite{Jitendra_Riesz_2024}, and phase transitions~\cite{Kethepalli_2022,Jitendra_Riesz_2024} can be studied in a controlled manner across different power-law exponents $k$.
% In recent years, there has been significant progress in understanding the collective and equilibrium properties of the RG, particularly in the large-$N$ limit~\cite{Riesz_avg_density_SA, Kethepalli_2021}. A series of works have elucidated the density profiles~\cite{Riesz_avg_density_SA}, edge fluctuations, and phase transitions in confined RG, 
These developments reveal rich universal structures beyond the traditional special limits of $k$. 
% For instance, exact density profiles in confined geometries have been obtained, exhibiting distinct regimes depending on the power exponent~\cite{Riesz_avg_density_SA}. Furthermore, studies of edge fluctuations have uncovered universal large deviation structures and higher-order phase transitions that persist across a wide class of long-range interactions~\cite{Kethepalli_2022}. 
Complementary investigations of stochastic dynamics have also explored fluctuation properties in overdamped settings, revealing anomalous scaling and connections to fractional Brownian motion~\cite{Kirone_2023_Riesz}. Together, these works establish the RG as a paradigmatic model for studying collective phenomena in power-law interacting systems.

Despite these advances, the understanding of real-time dynamics in the RG — both under Newtonian evolution or a collective formalism (such as hydrodynamics) remains unexplored. Most existing studies (with few exceptions~\cite{bks21}) have focused on equilibrium properties~\cite{Riesz_avg_density_SA,Saikat_Riesz_2022,Jitendra_Riesz_2024} or overdamped (Brownian) dynamics~\cite{Kirone_2023_Riesz}, leaving the deterministic Hamiltonian evolution largely unexplored. This gap is especially significant given that Newtonian dynamics provides the microscopic foundation for many-body systems and is essential for understanding transport, relaxation and approach to the Gibbs steady state.

An equally important and largely unexplored direction concerns the hydrodynamic description of the RG and its connection to microscopic dynamics. Hydrodynamics is expected to emerge as an effective theory governing the long-wavelength, low-frequency behavior of many-body systems~\cite{Forster1975,LandauLifshitzFluid}, provided there exists a separation of scales between microscopic dynamics and macroscopic evolution. While such a study has been carried out in a variety of classical and quantum systems, including gases~\cite{Subhadip_Blast_2021,Subhadip_Blast_2021_2, Singh2023, Amit2025_1,Amit2025Splash}, these have focused on the far from equilibrium regime. 
To our knowledge, a quantitative comparison between microscopic Newtonian dynamics and dissipative conventional hydrodynamics for a generically nonintegrable trapped many-body system, followed from strongly nonequilibrium initial conditions all the way to thermal equilibrium, has not been demonstrated. There has been much recent work on systems with weak integrability breaking where the framework of generalized hydrodynamics (GHD) has been used to address the question of thermalization to an eventual Gibbs state~\cite{Cao2018HardRods,AlvisePRL2020,alvise_review}.   

In this work, we address the important gap regarding the hydrodynamic description of a non-integrable Hamiltonian system and make a quantitative comparison  of the entire time evolution (right up to thermalization)   with microscopic Newtonian dynamics.  The RG provides a platform that is uniquely suited for such an investigation, since recent studies~\cite{Riesz_avg_density_SA,finite_range_Riesz_MK} have thoroughly characterized its thermodynamic properties. This  includes the precise form of the equation of state which is usually unavailable for interacting systems. The hydrodynamics of our system is described by two conserved fields (density and energy) and one non-conserved field (momentum). While the Euler-level equations are known exactly, at the level of Navier-Stokes, we introduce two phenomenological transport coefficients, namely the bulk viscosity and thermal conductivity.   
   Here we  provide  a comparison of the full time evolution of these three fields as extracted from hydrodynamics and Newtonian dynamics and find that there is a remarkable agreement. Thus  the approach to the Gibbs-Boltzmann state is quantitatively reproduced by a conventional three-field hydrodynamic theory over a broad range of timescales.

\textit{Model and properties.--} We consider a one-dimensional system of $N$ classical particles confined by a quadratic trap with pairwise power-law interaction, where each particle interacts with up to $d$ neighbors on either side~\cite{finite_range_Riesz_MK},
\begin{align}\label{eq:Eqntime_dynamics_pot}
     V_k ({\bf x})=\sum_{i=1}^N \frac{m\,\omega^2 x_i^2}{2}  +\frac{J}{2} \sum_{\substack{|i-j|\leq d \\ j\neq i}}^N   \frac{1}{|x_i-x_j|^k}.   
    \end{align}
Here $k>0$ is the power-law exponent, $J>0$ is the interaction strength and ${\bf x}=\{x_1,x_2,\ldots,x_N\}$, denotes the positions of the full set of particles.  All particles have mass $m$, and $\omega$ is the frequency of the harmonic trap. The parameter $d$ in
Eq.~\eqref{eq:Eqntime_dynamics_pot} restricts the interaction range. By tuning $d$ from $1$ to $N - 1,$ the model continuously interpolates between nearest neighbor coupling and all-to-all interactions. The model in Eq.~\eqref{eq:Eqntime_dynamics_pot} is a finite range generalization of the so-called Riesz gas for $d=N-1$ \cite{Riesz, Riesz_avg_density_SA}. We would like to emphasize that, in this paper, for the sake of convenience, we have considered a finite ranged model. However, our findings are valid for all-to-all coupling as long as we are in the relatively short-ranged regime which for all-to-all case corresponds to $k>1$.

\begin{figure*}[t]
\centering
\includegraphics[width=0.9\textwidth]{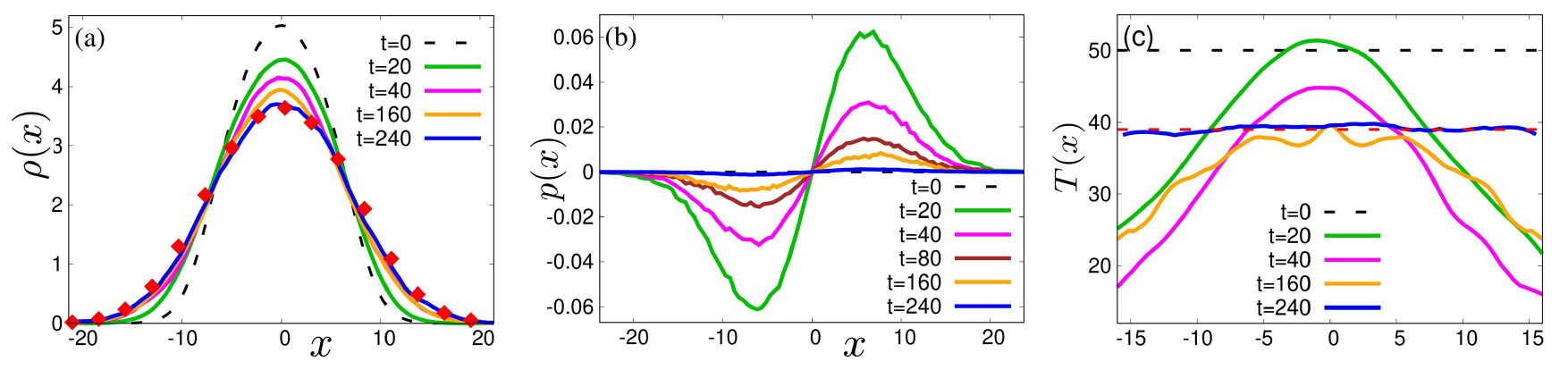}
\includegraphics[width=0.9\textwidth]{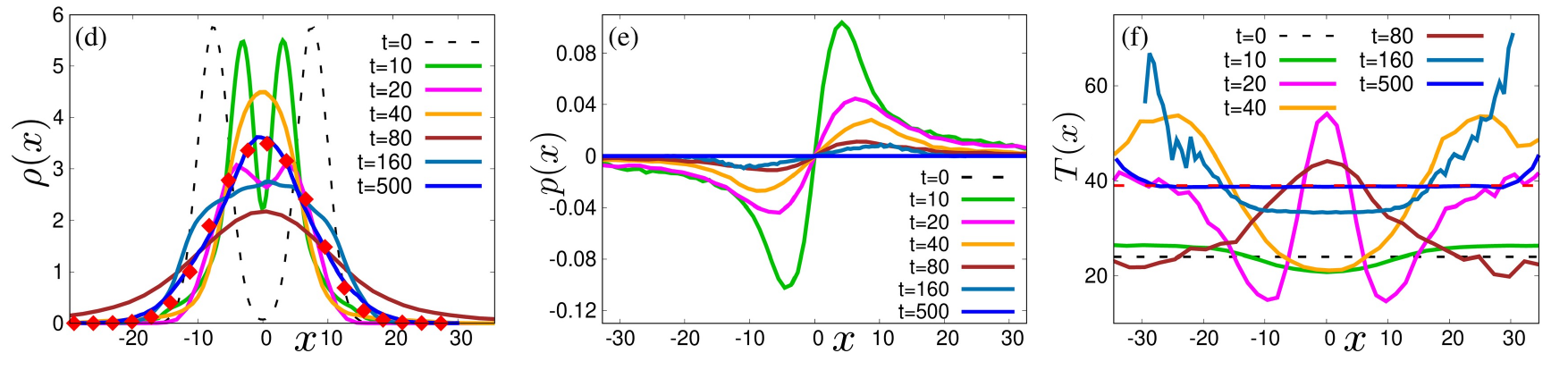}
\vspace*{-.2 cm}
\caption{(a), (b), (c) Newtonian dynamics results, following Eq.~\eqref{eq:Riesz_eom}, for the time evolution of density [Eq.~\eqref{eq:profile_rho_defn}], momentum [Eq.~\eqref{eq:profile_v_defn}] and temperature [Eq.~\eqref{eq:profile_T_defn}] fields, starting from the initial conditions corresponding to a dome-like initial profile (d), (e), (f) time evolution of density, momentum and temperature fields, starting from the initial conditions corresponding to a Newton-cradle-type initial profile. For both initial conditions, we consider $N=64$ and average over $10^4$ realizations. For (a), the steady state has been reached around $t\sim 240$, and for (d), the steady state has been reached around $t\sim 500$. The red squares in (a) and (d) are the expected steady-state density profiles obtained from Eq.~\eqref{eq:entropy_min} by rescaling the coordinate $x=yN^{\alpha_k}$ and  
and using the temperature $T\approx 39$ 
obtained using Eq.~\eqref{eq:virial}. 
The expected steady-state momentum profile in (b) and (e) is zero.  
}\label{Fig:rhoavg_micro}
\end{figure*}

For the energy given in Eq.~\eqref{eq:Eqntime_dynamics_pot}, the joint probability density function of the position of the particle at equilibrium becomes $P(x_1,x_2,\dots,x_N)=\exp(-\beta V_k [\{x_i\}])/Z_N(\beta)$ where $\beta$ is the inverse temperature (with $k_B=1$) and $Z_N(\beta)$ is the partition function. Following Eq.~\eqref{eq:Eqntime_dynamics_pot}, the harmonic trap tries to confine the particles near the center,  whereas the repulsive interactions drive them apart. The interplay of these competing effects leads to the fact that all the particles will settle down over a region  with a space-dependent average macroscopic density, 
$ \rho_N(x) =\frac{1}{N} \sum_{i=1}^N \langle \delta(x-x_i)\rangle$ (normalized to unity), where $\langle\dots\rangle$ denotes an average with respect to the Boltzmann weight. We note that the hydrodynamic density field will be defined as $\rho(x)=N \rho_N(x)$. 
%In the large $N$ limit, when $\beta\sim O(1)$, the equilibrium density profile for the energy in %Eq.~\eqref{eq:Eqntime_dynamics_pot} has been determined~\cite{supp} exactly for various interaction exponents $k$.
%over a support in the 
% range $[-\ell_k N^{\alpha_k}/2,\ell_k N^{\alpha_k}/2]$ (for $k\neq 1$) 
%with $\alpha_k=k/(k+2)$ for $k>1$ and $1/(k+2)$ for $-2<k<1$. 
%These results apply both to the fully connected case $(d=N-1)$ \cite{Riesz_avg_density_SA} and to interactions of finite range $d$ \cite{finite_range_Riesz_MK}. 

Let us define rescaled variables $y=xN^{-\alpha_k}$ and $\rho_k(y)=N^{\alpha_k} \rho_N(x)$. 
For  $\beta\sim O(N^{-2\alpha_k})$ where $\alpha_k = k/(k+2)$, one can then express the energy function $V_k(\{x_i\})$ in Eq.~\eqref{eq:Eqntime_dynamics_pot} as a functional of the density profile $\rho_k(y)$ in the following form \cite{Riesz_avg_density_SA,finite_range_Riesz_MK, Jitendra_Riesz_2024}
\begin{equation}\label{Riesz_action_Teff}
\begin{aligned}
\mathcal{V}_k[\rho_k(y)] ={}& \frac{1}{2}\int y^2\rho_k(y)\, dy + J \zeta_d(k)\int\rho_k(y)^{k+1}dy \\
& + T_{\rm eff}\int dy \,\rho_k(y)\log\rho_k(y) \\
& - \mu \left(\int \rho_k(y)\, dy -1\right), 
\end{aligned}
\end{equation}
and can write the coarse-grained partition function $Z_N(\beta) \sim \int d\mu \int \mathcal{D}[\rho_k(y)]~ {\rm exp}\left(-\beta N^{2\alpha_k+1} \mathcal{V}_k[\rho_k(y)]\right),$ with $\mu$ representing the Lagrange multiplier enforcing the
normalization of the density. We also defined  $\zeta_d(k)=\sum_{n=1}^d n^{-k}$, which is the truncated Riemann Zeta function and the effective temperature $T_{\rm eff}=T/N^{ \alpha_k} \sim O(1)$. Minimizing the action gives the saddle point equation 
\begin{align}\label{eq:entropy_min}
    \frac{y^2}{2}+J\zeta_d(k)(k+1)[\rho_k(y)]^k+T_{\rm{eff}} \log [\rho_k(y)]=\mu\,.
\end{align}
Solving Eq.~\eqref{eq:entropy_min}, we obtain the steady state density profile.  Note that at sufficiently low temperature $\beta >>N^{-2\alpha_k}$, one can neglect the entropy in the action or equivalently the last term in Eq.~\eqref{eq:entropy_min} and can obtain the exact density profile in this low temperature limit as mentioned in Ref. \cite{finite_range_Riesz_MK}. 

\textit{Newtonian evolution.--} We now add the kinetic energy term to Eq.~\eqref{eq:Eqntime_dynamics_pot} and the full Hamiltonian reads 
\begin{align}\label{eq:Eqntime_dynamics_1}
     H=\sum_{i=1}^N \frac{p_i^2}{2m} + V_k ({\bf x})  \end{align}
where $p_i$ denotes the momentum of the $i$-th particle and $V_k ({\bf x})$ is given in Eq.~\eqref{eq:Eqntime_dynamics_pot}. The resulting equations of motion are
\begin{gather}\label{eq:Riesz_eom}
    \dot x_i=p_i/m,~{\rm and} ~ \dot p_i=-m\omega^2x_i+\frac{Jk}{2}\sum_{\substack{|i-j|\leq d \\ j\neq i}}^N\frac{{\rm sgn}(x_i-x_j)}{|x_i-x_j|^{k+1}}.
\end{gather}
For a set of given initial conditions $\{x_i(0),p_i(0)\}$, Eq.~\eqref{eq:Riesz_eom} uniquely determines the subsequent time evolution of the system. We restrict our attention to the regime $d \sim O(1)$ and henceforth fix $d=1, k=1.5$, $m=1$ and $\omega=1$. We investigate the temporal evolution of the average density, momentum and energy profile and the eventual thermalization towards a steady state. 
%This analysis is carried out using both the microscopic dynamics generated by the equations of motion given in Eq.~\eqref{eq:Riesz_eom} and a complementary hydrodynamic description. In the steady state, the average density profiles obtained from both the microscopic dynamics and from the hydrodynamic description are expected to coincide and to agree with the profile $\rho_k(y)$ determined by the solution of the saddle-point equation in Eq.~\eqref{eq:entropy_min}.

\begin{figure*}[t]
\centering
\includegraphics[width=0.9\textwidth]{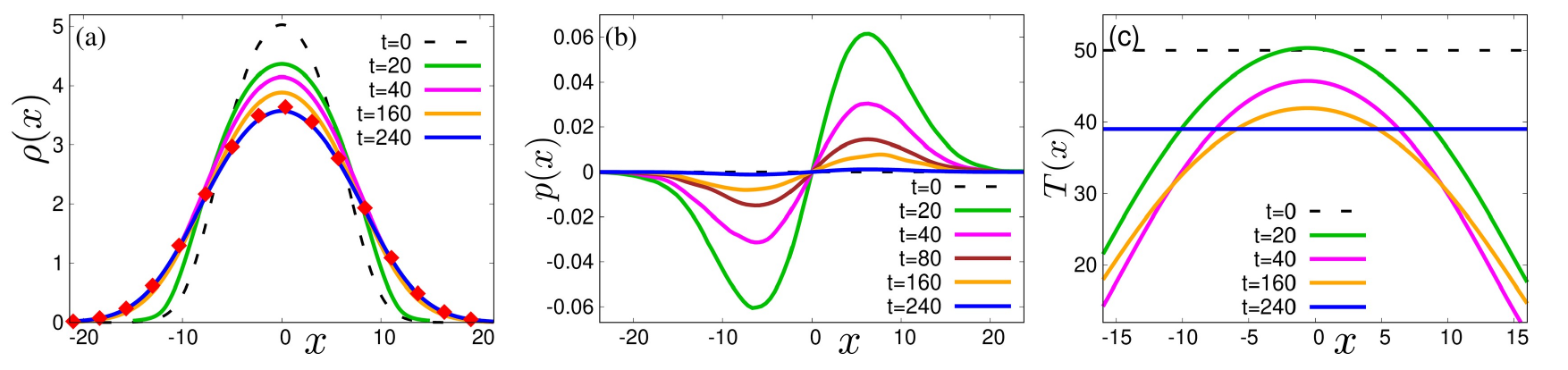}
\includegraphics[width=0.9\textwidth]{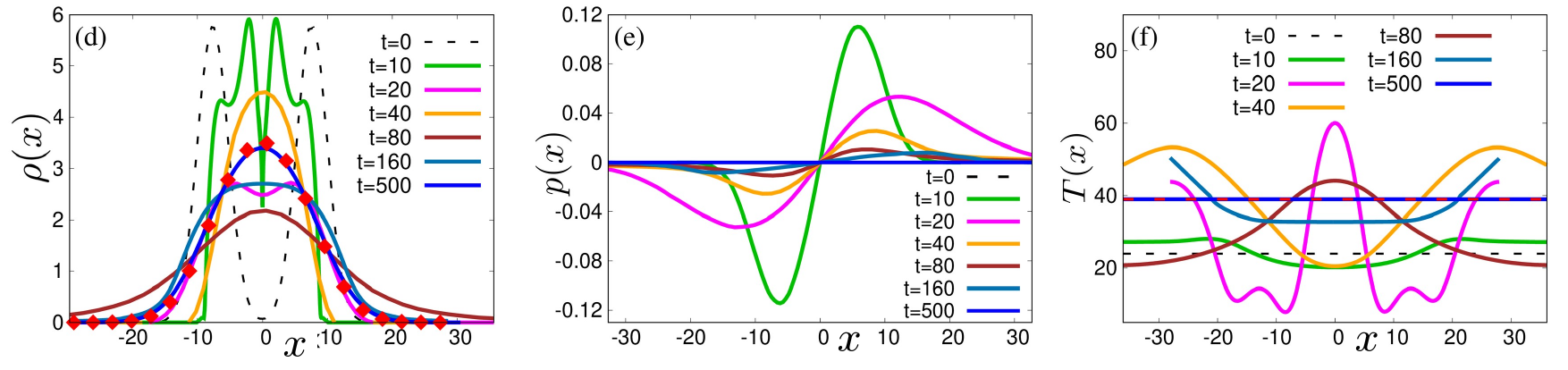}
\vspace*{-.2 cm}
\caption{  Evolution of the density,  momentum, and temperature profiles, obtained from the numerical simulation of the hydrodynamic equations in Eqs.~\eqref{eq:Hydro_eqns_density}-\eqref{eq:Hydro_eqns_energy} starting from the same initial condition as used in the Newtonian dynamics simulations of Fig. \ref{Fig:rhoavg_micro}. The initial conditions in (a) - (c) correspond to a dome-like initial profile, and (d) - (f) correspond to a Newton-cradle-type initial profile. Other parameters considered are $\xi=0.25, \kappa=0.15.$ For (a), the steady state has been reached around $t\sim 240$, and for (d), the steady state has been reached around $t\sim 500$, which is in agreement with the steady-state time scales obtained via Newtonian dynamics in Fig.~\ref{Fig:rhoavg_micro}. The red squares in (a) and (d) are the expected steady-state density profiles obtained from Eq.~\eqref{eq:entropy_min} by rescaling the coordinate $x=yN^{\alpha_k}$ and using the temperature $T\approx 39$ 
obtained from the hydrodynamics (see Eq.~(S30) in Supplemental Material~\cite{supp}). The expected steady-state momentum profile in (b) and (e) is zero. 
%The steady-state temperature obtained in (c) and (f) is $T\sim 39$. 
}
\label{Fig:rhoavg_hydro}
\end{figure*}

%\begin{figure}
%\centering
%\includegraphics[width=4.27cm]{Main_Plots/rho_dome_hydroa.eps}
%\includegraphics[width=4.27cm]{Main_Plots/v_dome_hydro.eps}
%\includegraphics[width=4.27cm]{Main_Plots/rho_NC_hydroa.eps}
%\includegraphics[width=4.27cm]{Main_Plots/v_NC_hydro.eps}
%\vspace*{-.2 cm}
%\caption{  Evolution of the density and velocity profiles, obtained from the numerical simulation of the hydrodynamic equations in Eqs.~\eqref{eq:Hydro_eqns_density}-\eqref{eq:Hydro_eqns_energy} starting from the same initial condition as used in the Newtonian dynamics simulations of Fig. \ref{Fig:rhoavg_micro}. The initial conditions in (a) - (b) correspond to a Dome-like initial profile, and (c) - (d) correspond to a Newton-cradle-type initial profile. Other parameters considered as $\xi=0.25, \kappa=0.15.$ For (a), the steady state has reached around $t\sim 240$, and for (b), the steady state has reached around $t\sim 500$, which is in agreement with the steady state time scales obtained via Newtonian dynamics in Fig.~\ref{Fig:rhoavg_micro}. The red squares in (a) and (c) are the expected steady-state density profiles obtained from Eq.~\eqref{eq:entropy_min}. The expected steady state velocity profile in (b) and (d) is zero.}\label{Fig:rhoavg_hydro}
%\end{figure}

We simulate Eq.~\eqref{eq:Riesz_eom} employing standard molecular dynamics based on a symplectic velocity-Verlet integration scheme~\cite{Swope1982}, with a sufficiently small time step $dt=10^{-6}$ to ensure numerical accuracy. To study dynamics and approach to the steady state, we initialize the positions of all particles $\{x_i(0)\}$, $i=1,2,\ldots,N$,  from prescribed spatial density profiles $\rho(x)$.  The velocities $\{v_i(0)\}$ are first chosen from a Maxwell distributions at some temperature and then suitably rescaled such that the total energy of the system has a fixed prescribed value. This defines our initial ensemble which is a subset of the constant energy surface.  
In particular, we consider two distinct choices for the initial density profiles: (i) a dome-like profile drawn from an equilibrium distribution at a chosen temperature \cite{Riesz_avg_density_SA}, and (ii) a Newton-cradle-type profile~\cite{Kinoshita2006} consisting of two blobs, each of finite width and separated by a finite distance. The details of the microscopic simulations and the procedure used to extract the steady-state density profile are provided in the Supplemental Material~\cite{supp}.   
For both the initial conditions, the steady-state temperature is determined using the generalized equipartition relation \cite{virial}, 
\begin{align}
\langle p_i^2\rangle = \Big\langle x_i \frac{\partial H}{\partial x_i}\Big\rangle =k_B T.
\label{eq:virial}
\end{align}
In the simulations, the ensemble average is obtained by averaging over the steady state. At this temperature, the density profile predicted by using the analytic thermodynamic result [Eq.~\eqref{eq:entropy_min}] is expected to coincide with the equilibrium density profile obtained from Newtonian dynamics [Eq.~\eqref{eq:Riesz_eom}].

It is common in hydrodynamics to work with the more physical velocity and  temperature fields instead of the momentum and energy fields.  
These fields are constructed from $\{x_i(t),v_i(t)\}$ by 
\begin{gather}
    \rho(x,t)=\Big \langle \sum_{i=1}^N\delta(x-x_i(t))\Big\rangle \label{eq:profile_rho_defn}\\
    p(x,t)=\rho(x,t)v(x,t)= \Big \langle \sum_{i=1}^N v_i(t)~\delta(x-x_i(t))\Big\rangle \label{eq:profile_v_defn}\\
    \rho(x,t) T(x,t)=\Big \langle \sum_{i=1}^N (v_i(t)-v(x,t))^2~\delta(x-x_i(t))\Big\rangle\label{eq:profile_T_defn},
\end{gather}
where  $\langle\cdots\rangle$ denotes the average over  initial conditions drawn from an initial ensemble (see above).   

Fig.~\ref{Fig:rhoavg_micro} shows the results of the microscopic simulations for the time evolution of the average density $\rho(x,t)$, momentum   $p(x,t)$, and temperature $T(x,t)$ profiles, together with a comparison to the expected steady-state density profile obtained from Eq.~\eqref{eq:entropy_min} by rescaling the coordinate $x=yN^{\alpha_k}$, and using the steady state temperature value, $T\approx 39$, obtained from Eq.~\eqref{eq:virial}. Since momentum is not conserved, the expected steady-state momentum density vanishes. 

\textit{Hydrodynamic Equations.--} In the absence of the trapping potenital $U(x)$, we have three hydrodynamic fields, $\rho,p,e$, corresponding to mass, momentum and energy conservation. The corresponding hydrodynamic equations are well known~\cite{LandauLifshitzFluid,Subhadip_Blast_2021,Subhadip_Blast_2021_2}. The presence of the trap destroys momentum conservation but one can still formulate a hydrodynamic description in terms of the conserved fields $\rho,e$ which are coupled to the slowly decaying momentum field, $p$. The momentum equation gets an extra term related to the external force, $-\partial_x U$, and the resulting hydrodynamic equations are~\cite{supp}
\begin{eqnarray}
\label{eq:Hydro_eqns1}
\partial_{t}\rho+\partial_x(\rho v)=0\\
\label{eq:Hydro_eqns2}
\partial_{t}(\rho v)+\partial_x(\rho v^2+P)+\rho\partial_xU(x)=\partial_x(\xi\partial_x v)\\
\label{eq:Hydro_eqns3}
\partial_{t}(\rho e)+\partial_x(\rho v e+Pv)=\partial_x(\xi v \partial_x v+\kappa\partial_xT),
\end{eqnarray}
where $\xi$ denotes the bulk viscosity and $\kappa$ is the thermal conductivity of the system.
Here $e=\epsilon(x,t)+v^2/2$, where $\epsilon(x,t)$ denotes the internal energy per unit mass~\cite{supp}, and $P$ is the pressure, following the Riesz gas equation of state. For the finite-range Riesz gas at temperature $T$, the explicit form of pressure $P$ as a function of the density field $\rho(x,t)$ can be expressed as~\cite{supp}
\begin{align}\label{eq:Hydro_eqns_P}
    P(x,t)=J \zeta_d(k) k ~\rho(x,t)^{k+1}+T(x,t)\rho(x,t).
\end{align}
Substituting Eq.~\eqref{eq:Hydro_eqns_P} into Eqs.~\eqref{eq:Hydro_eqns2}-\eqref{eq:Hydro_eqns3}, and using the relation between internal energy, $\epsilon(x,t)$ and temperature, $T(x,t)$ derived in the Supplemental Material~\cite{supp}, and performing straightforward algebra, yields the hydrodynamic equations
\begin{gather}
\partial_t\rho+\partial_x(\rho v)=0\label{eq:Hydro_eqns_density}\\
    \rho( \partial_t +  v\partial_x )v+\partial_x(J\zeta_d(k)k\rho^{k+1}  + \rho T)\cr
    \qquad\qquad\qquad\qquad+\rho\partial_xU =\partial_x(\xi \partial_xv)\label{eq:Hydro_eqns_momentum}\\
     \frac{\rho^3}{2}\Big(\partial_t+v\partial_x\Big)\Bigg(\frac{T}{\rho^2}\Bigg)=\xi  (\partial_x v)^2+\partial_x (\kappa\partial_x T) \label{eq:Hydro_eqns_energy}
\end{gather}
We solve Eqs.~\eqref{eq:Hydro_eqns_density}-\eqref{eq:Hydro_eqns_energy} numerically. The details of the numerical simulation are discussed in the supplemental material~\cite{supp}.
Fig.~\ref{Fig:rhoavg_hydro} displays the hydrodynamic results for the time evolution of the density $\rho(x,t)$, momentum $p(x,t)$, and temperature $T(x,t)$ profiles. The steady-state density obtained, from Eq.~\eqref{eq:entropy_min}, by rescaling the coordinate $x=yN^{\alpha_k}$, is shown for comparison. In the steady state, the momentum field is expected to be zero.

\begin{figure*}[t]
\centering
\includegraphics[width=0.9\textwidth]{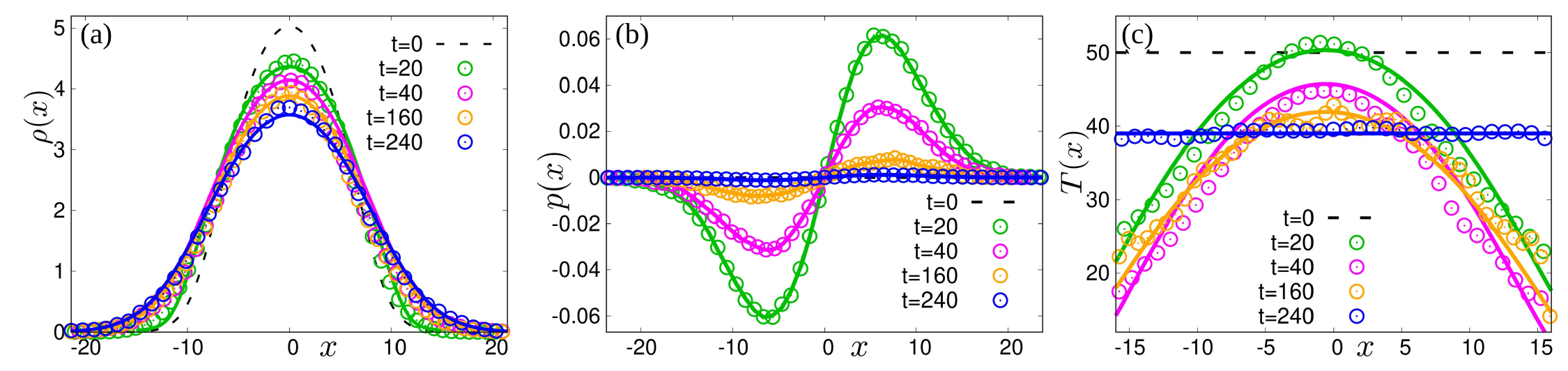}
\includegraphics[width=0.9\textwidth]{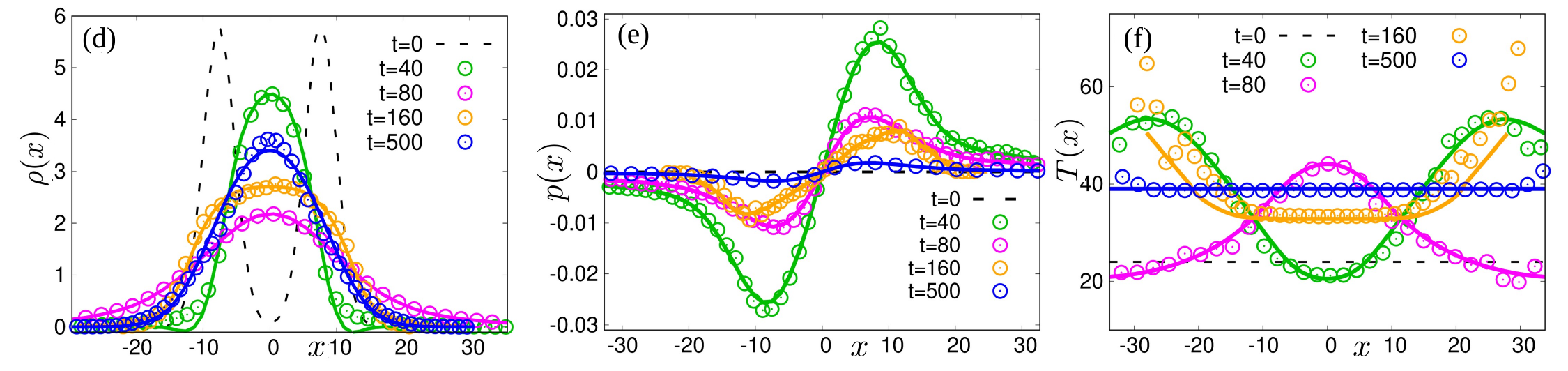}
\vspace*{-.2 cm}
\caption{Comparison of the density, momentum, and temperature profiles at different time steps, obtained from the Newtonian dynamics following Eq.~\eqref{eq:Riesz_eom} (circles) with the numerical simulation of the hydrodynamic equations in Eqs.~\eqref{eq:Hydro_eqns_density}-\eqref{eq:Hydro_eqns_energy} (solid lines). Panels (a)-(c) show results for a dome-like initial profile, while panels (d)-(f) show results for a Newton-cradle-type initial profile, with $\xi=0.25, \kappa=0.15$. All other parameters for the  Newtonian dynamics and the hydrodynamic simulations are the same as in Figs.~\ref{Fig:rhoavg_micro} and \ref{Fig:rhoavg_hydro}, respectively.
}\label{Fig:rho_comp_micro_hydro}
\end{figure*}

\textit{Comparison between Newtonian dynamics and hydrodynamics.--}
We now compare, at a quantitative level, the results obtained via Newtonian dynamics [Eq,~\eqref{eq:Riesz_eom}] and hydrodynamics [Eqs.~\eqref{eq:Hydro_eqns_density}-\eqref{eq:Hydro_eqns_energy}]. We start with the same initial condition for both the Newtonian dynamics and hydrodynamics simulations. In Fig.~\ref{Fig:rho_comp_micro_hydro}, we show this comparison (for density, momentum, and temperature) for a significant window of time scales, starting from early all the way till the time scales at which the steady state is reached. We demonstrate excellent agreement between Newtonian dynamics and hydrodynamic simulations for both the dome-like and Newton's-cradle-like initial conditions. The Newton's-cradle-like initial condition is fairly extreme in nature, i.e., very far from the final steady state. Nonetheless, the hydrodynamic simulations show excellent agreement with Newtonian dynamics. For the Newton's-cradle-like initial condition, we have omitted showing very small times, as these can be more sensitive to the precise choice of the bulk viscosity $\xi$ and thermal conductivity $\kappa$. However, for the rest of the time-scales our results are robust~\cite{supp} to the exact value of $\xi$ and $\kappa$. In other words, we find that there is a considerable window in $\xi$ and $\kappa$ where we find excellent quantitative agreement. We also note that although, we focus on $k=1.5$ and $d=1$, our results hold for any $k>0$ and $d\sim O(1)$ where the information of $k$ and $d$ is encoded in Eq.~\eqref{eq:Riesz_eom} for Newtonian dynamics and in Eq.~\eqref{eq:Hydro_eqns_P} for hydrodynamics (with no phenomenological parameter).

\textit{Conclusions and Outlook.--}
In this work, we have demonstrated that the real-time dynamics of the Riesz gas can be quantitatively captured by a hydrodynamic framework built on coarse-grained fields of density, velocity, and temperature. By systematically comparing large-scale Newtonian simulations (Fig.~\ref{Fig:rhoavg_micro}) with hydrodynamic evolution (Fig.~\ref{Fig:rhoavg_hydro}) across representative initial conditions, we established a clear and robust correspondence (Fig.~\ref{Fig:rho_comp_micro_hydro}) between microscopic and macroscopic descriptions. The inclusion of phenomenological transport coefficients, namely bulk viscosity and thermal conductivity, proved sufficient to encode dissipative effects and reproduce the observed dynamics over a wide range of timescales all the way till the system thermalizes (see Fig.~\ref{Fig:rho_comp_micro_hydro}). Importantly, the persistence of agreement across a broad parameter window highlights the emergent nature and robustness of the hydrodynamic description --- in particular it demonstrates that the nonequilibrium evolution to the steady state is predominantly controlled by the Euler dynamics. Our results thus provide a concrete and first-principles validation of hydrodynamics in the Riesz gas, for a system starting far from equilibrium and eventually thermalizing.

The present study opens several promising directions for future investigation. A natural important direction is the microscopic derivation of the transport coefficients, which would deepen the connection between particle-level dynamics and emergent dissipation. It would also be of interest to explore regimes where hydrodynamics may break down, such as extreme initial states and finite-size effects. It will be interesting to explore certain special integrable points (such as the Calogero-Moser case of $k=2$~\cite{Calogero_JMP_1969,Calogero_JMP_1971,Calogero_LNC_1975,OLSHANETSKY1981, polychronakos_IOP_2006} with all-to-all coupling) where conventional hydrodynamics needs to be replaced by generalized hydrodynamics \cite{Doyon_GHD_PRX,Jacopo_GHD_2016,ESSLER2023GHD,Doyon_GHD_PRX2025,Cao2018HardRods,AlvisePRL2020,alvise_review} due to classical integrability of the model. In this context, it would also be interesting to understand how weak perturbations away from integrability lead to a crossover from generalized to conventional hydrodynamics, as explored in recent works \cite{santra2026crossoverGHD}. Finally, given the growing experimental accessibility of systems~\cite{James1998,brown_CUP_2003,chalony_PRA_2013,zhang_nature_2017} with tunable power-law interactions, our framework provides a foundation for connecting theoretical predictions with experiments, potentially enabling direct probes of emergent hydrodynamics in power-law interacting systems.

\textit{Acknowledgments.--}
We thank A. Kundu, J. Kethepalli and R. Mukherjee for useful discussions. AD acknowledges support from the J.C.~Bose Fellowship (JCB/2022/000014) of the Science and Engineering Research Board, Department of Science and Technology, Government of India. We also acknowledge support from the Department of Atomic Energy, Government of India, under Project No. RTI4001. 

\bibliography{references}

\begin{thebibliography}{49}%
\makeatletter
\providecommand \@ifxundefined [1]{%
 \@ifx{#1\undefined}
}%
\providecommand \@ifnum [1]{%
 \ifnum #1\expandafter \@firstoftwo
 \else \expandafter \@secondoftwo
 \fi
}%
\providecommand \@ifx [1]{%
 \ifx #1\expandafter \@firstoftwo
 \else \expandafter \@secondoftwo
 \fi
}%
\providecommand \natexlab [1]{#1}%
\providecommand \enquote  [1]{``#1''}%
\providecommand \bibnamefont  [1]{#1}%
\providecommand \bibfnamefont [1]{#1}%
\providecommand \citenamefont [1]{#1}%
\providecommand \href@noop [0]{\@secondoftwo}%
\providecommand \href [0]{\begingroup \@sanitize@url \@href}%
\providecommand \@href[1]{\@@startlink{#1}\@@href}%
\providecommand \@@href[1]{\endgroup#1\@@endlink}%
\providecommand \@sanitize@url [0]{\catcode `\\12\catcode `\$12\catcode `\&12\catcode `\#12\catcode `\^12\catcode `\_12\catcode `\%12\relax}%
\providecommand \@@startlink[1]{}%
\providecommand \@@endlink[0]{}%
\providecommand \url  [0]{\begingroup\@sanitize@url \@url }%
\providecommand \@url [1]{\endgroup\@href {#1}{\urlprefix }}%
\providecommand \urlprefix  [0]{URL }%
\providecommand \Eprint [0]{\href }%
\providecommand \doibase [0]{https://doi.org/}%
\providecommand \selectlanguage [0]{\@gobble}%
\providecommand \bibinfo  [0]{\@secondoftwo}%
\providecommand \bibfield  [0]{\@secondoftwo}%
\providecommand \translation [1]{[#1]}%
\providecommand \BibitemOpen [0]{}%
\providecommand \bibitemStop [0]{}%
\providecommand \bibitemNoStop [0]{.\EOS\space}%
\providecommand \EOS [0]{\spacefactor3000\relax}%
\providecommand \BibitemShut  [1]{\csname bibitem#1\endcsname}%
\let\auto@bib@innerbib\@empty
%</preamble>
\bibitem [{\citenamefont {Riesz}(1938)}]{Riesz}%
  \BibitemOpen
  \bibfield  {author} {\bibinfo {author} {\bibfnamefont {M.}~\bibnamefont {Riesz}},\ }\bibfield  {title} {\bibinfo {title} {Intégrales de riemann–liouville et potentiels},\ }\href {https://www.acta.bibl.u-szeged.hu/13487/1/math_009_001-042.pdf} {\bibfield  {journal} {\bibinfo  {journal} {Acta Sci. Math. Univ. Szeged}\ }\textbf {\bibinfo {volume} {9}},\ \bibinfo {pages} {1} (\bibinfo {year} {1938})}\BibitemShut {NoStop}%
\bibitem [{\citenamefont {Lewin}(2022)}]{Lewin_Riesz}%
  \BibitemOpen
  \bibfield  {author} {\bibinfo {author} {\bibfnamefont {M.}~\bibnamefont {Lewin}},\ }\bibfield  {title} {\bibinfo {title} {Coulomb and riesz gases: The known and the unknown},\ }\href {https://doi.org/10.1063/5.0086835} {\bibfield  {journal} {\bibinfo  {journal} {J. Math. Phys.}\ }\textbf {\bibinfo {volume} {63}},\ \bibinfo {pages} {061101} (\bibinfo {year} {2022})}\BibitemShut {NoStop}%
\bibitem [{\citenamefont {Agarwal}\ \emph {et~al.}(2019)\citenamefont {Agarwal}, \citenamefont {Dhar}, \citenamefont {Kulkarni}, \citenamefont {Kundu}, \citenamefont {Majumdar}, \citenamefont {Mukamel},\ and\ \citenamefont {Schehr}}]{Riesz_avg_density_SA}%
  \BibitemOpen
  \bibfield  {author} {\bibinfo {author} {\bibfnamefont {S.}~\bibnamefont {Agarwal}}, \bibinfo {author} {\bibfnamefont {A.}~\bibnamefont {Dhar}}, \bibinfo {author} {\bibfnamefont {M.}~\bibnamefont {Kulkarni}}, \bibinfo {author} {\bibfnamefont {A.}~\bibnamefont {Kundu}}, \bibinfo {author} {\bibfnamefont {S.~N.}\ \bibnamefont {Majumdar}}, \bibinfo {author} {\bibfnamefont {D.}~\bibnamefont {Mukamel}},\ and\ \bibinfo {author} {\bibfnamefont {G.}~\bibnamefont {Schehr}},\ }\bibfield  {title} {\bibinfo {title} {Harmonically confined particles with long-range repulsive interactions},\ }\href {https://doi.org/10.1103/PhysRevLett.123.100603} {\bibfield  {journal} {\bibinfo  {journal} {Phys. Rev. Lett.}\ }\textbf {\bibinfo {volume} {123}},\ \bibinfo {pages} {100603} (\bibinfo {year} {2019})}\BibitemShut {NoStop}%
\bibitem [{\citenamefont {{Lenard}}(1963)}]{Lenard_JMP_1963}%
  \BibitemOpen
  \bibfield  {author} {\bibinfo {author} {\bibfnamefont {A.}~\bibnamefont {{Lenard}}},\ }\bibfield  {title} {\bibinfo {title} {{Exact Statistical Mechanics of a One-Dimensional System with Coulomb Forces. III. Statistics of the Electric Field}},\ }\href {https://doi.org/10.1063/1.1703988} {\bibfield  {journal} {\bibinfo  {journal} {Journal of Mathematical Physics}\ }\textbf {\bibinfo {volume} {4}},\ \bibinfo {pages} {533} (\bibinfo {year} {1963})}\BibitemShut {NoStop}%
\bibitem [{\citenamefont {Baxter}(1963)}]{Statistical_MPCPS_Baxter}%
  \BibitemOpen
  \bibfield  {author} {\bibinfo {author} {\bibfnamefont {R.~J.}\ \bibnamefont {Baxter}},\ }\bibfield  {title} {\bibinfo {title} {Statistical mechanics of a one-dimensional coulomb system with a uniform charge background},\ }\href {https://doi.org/10.1017/S0305004100003790} {\bibfield  {journal} {\bibinfo  {journal} {Mathematical Proceedings of the Cambridge Philosophical Society}\ }\textbf {\bibinfo {volume} {59}},\ \bibinfo {pages} {779–787} (\bibinfo {year} {1963})}\BibitemShut {NoStop}%
\bibitem [{\citenamefont {Dhar}\ \emph {et~al.}(2018)\citenamefont {Dhar}, \citenamefont {Kundu}, \citenamefont {Majumdar}, \citenamefont {Sabhapandit},\ and\ \citenamefont {Schehr}}]{dhar_IOP_2018}%
  \BibitemOpen
  \bibfield  {author} {\bibinfo {author} {\bibfnamefont {A.}~\bibnamefont {Dhar}}, \bibinfo {author} {\bibfnamefont {A.}~\bibnamefont {Kundu}}, \bibinfo {author} {\bibfnamefont {S.~N.}\ \bibnamefont {Majumdar}}, \bibinfo {author} {\bibfnamefont {S.}~\bibnamefont {Sabhapandit}},\ and\ \bibinfo {author} {\bibfnamefont {G.}~\bibnamefont {Schehr}},\ }\bibfield  {title} {\bibinfo {title} {Extreme statistics and index distribution in the classical 1dcoulomb gas},\ }\href {https://doi.org/10.1088/1751-8121/aac75f} {\bibfield  {journal} {\bibinfo  {journal} {Journal of Physics A: Mathematical and Theoretical}\ }\textbf {\bibinfo {volume} {51}},\ \bibinfo {pages} {295001} (\bibinfo {year} {2018})}\BibitemShut {NoStop}%
\bibitem [{\citenamefont {Dhar}\ \emph {et~al.}(2017)\citenamefont {Dhar}, \citenamefont {Kundu}, \citenamefont {Majumdar}, \citenamefont {Sabhapandit},\ and\ \citenamefont {Schehr}}]{dhar_PRL_2017}%
  \BibitemOpen
  \bibfield  {author} {\bibinfo {author} {\bibfnamefont {A.}~\bibnamefont {Dhar}}, \bibinfo {author} {\bibfnamefont {A.}~\bibnamefont {Kundu}}, \bibinfo {author} {\bibfnamefont {S.~N.}\ \bibnamefont {Majumdar}}, \bibinfo {author} {\bibfnamefont {S.}~\bibnamefont {Sabhapandit}},\ and\ \bibinfo {author} {\bibfnamefont {G.}~\bibnamefont {Schehr}},\ }\bibfield  {title} {\bibinfo {title} {Exact extremal statistics in the classical 1d coulomb gas},\ }\href {https://doi.org/10.1103/PhysRevLett.119.060601} {\bibfield  {journal} {\bibinfo  {journal} {Phys. Rev. Lett.}\ }\textbf {\bibinfo {volume} {119}},\ \bibinfo {pages} {060601} (\bibinfo {year} {2017})}\BibitemShut {NoStop}%
\bibitem [{\citenamefont {{Dyson}}(1962)}]{Dyson_JMP_1962}%
  \BibitemOpen
  \bibfield  {author} {\bibinfo {author} {\bibfnamefont {F.~J.}\ \bibnamefont {{Dyson}}},\ }\bibfield  {title} {\bibinfo {title} {{Statistical Theory of the Energy Levels of Complex Systems. I}},\ }\href {https://doi.org/10.1063/1.1703773} {\bibfield  {journal} {\bibinfo  {journal} {Journal of Mathematical Physics}\ }\textbf {\bibinfo {volume} {3}},\ \bibinfo {pages} {140} (\bibinfo {year} {1962})}\BibitemShut {NoStop}%
\bibitem [{\citenamefont {Dyson}(1962)}]{dyson_JKP_1962}%
  \BibitemOpen
  \bibfield  {author} {\bibinfo {author} {\bibfnamefont {F.~J.}\ \bibnamefont {Dyson}},\ }\bibfield  {title} {\bibinfo {title} {Statistical theory of the energy levels of complex systems. ii},\ }\href {https://doi.org/10.1063/1.1703774} {\bibfield  {journal} {\bibinfo  {journal} {Journal of Mathematical Physics}\ }\textbf {\bibinfo {volume} {3}},\ \bibinfo {pages} {157} (\bibinfo {year} {1962})}\BibitemShut {NoStop}%
\bibitem [{\citenamefont {Mehta}(2004)}]{mehta2004random}%
  \BibitemOpen
  \bibfield  {author} {\bibinfo {author} {\bibfnamefont {M.~L.}\ \bibnamefont {Mehta}},\ }\href@noop {} {\emph {\bibinfo {title} {Random matrices}}},\ Vol.\ \bibinfo {volume} {142}\ (\bibinfo  {publisher} {Elsevier},\ \bibinfo {year} {2004})\BibitemShut {NoStop}%
\bibitem [{\citenamefont {Forrester}(2010)}]{forrester2010log}%
  \BibitemOpen
  \bibfield  {author} {\bibinfo {author} {\bibfnamefont {P.~J.}\ \bibnamefont {Forrester}},\ }\href@noop {} {\emph {\bibinfo {title} {Log-gases and random matrices (LMS-34)}}}\ (\bibinfo  {publisher} {Princeton university press},\ \bibinfo {year} {2010})\BibitemShut {NoStop}%
\bibitem [{\citenamefont {{Calogero}}(1969)}]{Calogero_JMP_1969}%
  \BibitemOpen
  \bibfield  {author} {\bibinfo {author} {\bibfnamefont {F.}~\bibnamefont {{Calogero}}},\ }\bibfield  {title} {\bibinfo {title} {{Ground State of a One-Dimensional N-Body System}},\ }\href {https://doi.org/10.1063/1.1664821} {\bibfield  {journal} {\bibinfo  {journal} {Journal of Mathematical Physics}\ }\textbf {\bibinfo {volume} {10}},\ \bibinfo {pages} {2197} (\bibinfo {year} {1969})}\BibitemShut {NoStop}%
\bibitem [{\citenamefont {{Calogero}}(1971)}]{Calogero_JMP_1971}%
  \BibitemOpen
  \bibfield  {author} {\bibinfo {author} {\bibfnamefont {F.}~\bibnamefont {{Calogero}}},\ }\bibfield  {title} {\bibinfo {title} {{Solution of the One-Dimensional N-Body Problems with Quadratic and/or Inversely Quadratic Pair Potentials}},\ }\href {https://doi.org/10.1063/1.1665604} {\bibfield  {journal} {\bibinfo  {journal} {Journal of Mathematical Physics}\ }\textbf {\bibinfo {volume} {12}},\ \bibinfo {pages} {419} (\bibinfo {year} {1971})}\BibitemShut {NoStop}%
\bibitem [{\citenamefont {Calogero}(1975)}]{Calogero_LNC_1975}%
  \BibitemOpen
  \bibfield  {author} {\bibinfo {author} {\bibfnamefont {F.}~\bibnamefont {Calogero}},\ }\bibfield  {title} {\bibinfo {title} {{Exactly Solvable One-Dimensional Many Body Problems}},\ }\href {https://doi.org/10.1007/BF02790495} {\bibfield  {journal} {\bibinfo  {journal} {Lett. Nuovo Cim.}\ }\textbf {\bibinfo {volume} {13}},\ \bibinfo {pages} {411} (\bibinfo {year} {1975})}\BibitemShut {NoStop}%
\bibitem [{\citenamefont {Olshanetsky}\ and\ \citenamefont {Perelomov}(1981)}]{OLSHANETSKY1981}%
  \BibitemOpen
  \bibfield  {author} {\bibinfo {author} {\bibfnamefont {M.}~\bibnamefont {Olshanetsky}}\ and\ \bibinfo {author} {\bibfnamefont {A.}~\bibnamefont {Perelomov}},\ }\bibfield  {title} {\bibinfo {title} {Classical integrable finite-dimensional systems related to lie algebras},\ }\href {https://doi.org/https://doi.org/10.1016/0370-1573(81)90023-5} {\bibfield  {journal} {\bibinfo  {journal} {Physics Reports}\ }\textbf {\bibinfo {volume} {71}},\ \bibinfo {pages} {313} (\bibinfo {year} {1981})}\BibitemShut {NoStop}%
\bibitem [{\citenamefont {Polychronakos}(2006)}]{polychronakos_IOP_2006}%
  \BibitemOpen
  \bibfield  {author} {\bibinfo {author} {\bibfnamefont {A.~P.}\ \bibnamefont {Polychronakos}},\ }\bibfield  {title} {\bibinfo {title} {The physics and mathematics of calogero particles},\ }\href {https://doi.org/10.1088/0305-4470/39/41/s07} {\bibfield  {journal} {\bibinfo  {journal} {Journal of Physics A: Mathematical and General}\ }\textbf {\bibinfo {volume} {39}},\ \bibinfo {pages} {12793–12845} (\bibinfo {year} {2006})}\BibitemShut {NoStop}%
\bibitem [{\citenamefont {Tonks}(1936)}]{HR_Tonks}%
  \BibitemOpen
  \bibfield  {author} {\bibinfo {author} {\bibfnamefont {L.}~\bibnamefont {Tonks}},\ }\bibfield  {title} {\bibinfo {title} {The complete equation of state of one, two and three-dimensional gases of hard elastic spheres},\ }\href {https://doi.org/10.1103/PhysRev.50.955} {\bibfield  {journal} {\bibinfo  {journal} {Phys. Rev.}\ }\textbf {\bibinfo {volume} {50}},\ \bibinfo {pages} {955} (\bibinfo {year} {1936})}\BibitemShut {NoStop}%
\bibitem [{\citenamefont {Kethepalli}\ \emph {et~al.}(2023)\citenamefont {Kethepalli}, \citenamefont {Bagchi}, \citenamefont {Dhar}, \citenamefont {Kulkarni},\ and\ \citenamefont {Kundu}}]{JK_HardRods}%
  \BibitemOpen
  \bibfield  {author} {\bibinfo {author} {\bibfnamefont {J.}~\bibnamefont {Kethepalli}}, \bibinfo {author} {\bibfnamefont {D.}~\bibnamefont {Bagchi}}, \bibinfo {author} {\bibfnamefont {A.}~\bibnamefont {Dhar}}, \bibinfo {author} {\bibfnamefont {M.}~\bibnamefont {Kulkarni}},\ and\ \bibinfo {author} {\bibfnamefont {A.}~\bibnamefont {Kundu}},\ }\bibfield  {title} {\bibinfo {title} {Finite-temperature equilibrium density profiles of integrable systems in confining potentials},\ }\href {https://doi.org/10.1103/PhysRevE.107.044101} {\bibfield  {journal} {\bibinfo  {journal} {Phys. Rev. E}\ }\textbf {\bibinfo {volume} {107}},\ \bibinfo {pages} {044101} (\bibinfo {year} {2023})}\BibitemShut {NoStop}%
\bibitem [{\citenamefont {Kumar}\ \emph {et~al.}(2020)\citenamefont {Kumar}, \citenamefont {Kulkarni},\ and\ \citenamefont {Kundu}}]{finite_range_Riesz_MK}%
  \BibitemOpen
  \bibfield  {author} {\bibinfo {author} {\bibfnamefont {A.}~\bibnamefont {Kumar}}, \bibinfo {author} {\bibfnamefont {M.}~\bibnamefont {Kulkarni}},\ and\ \bibinfo {author} {\bibfnamefont {A.}~\bibnamefont {Kundu}},\ }\bibfield  {title} {\bibinfo {title} {Particles confined in arbitrary potentials with a class of finite-range repulsive interactions},\ }\href {https://doi.org/10.1103/PhysRevE.102.032128} {\bibfield  {journal} {\bibinfo  {journal} {Phys. Rev. E}\ }\textbf {\bibinfo {volume} {102}},\ \bibinfo {pages} {032128} (\bibinfo {year} {2020})}\BibitemShut {NoStop}%
\bibitem [{\citenamefont {James}(1998)}]{James1998}%
  \BibitemOpen
  \bibfield  {author} {\bibinfo {author} {\bibfnamefont {D.~F.~V.}\ \bibnamefont {James}},\ }\bibfield  {title} {\bibinfo {title} {Quantum dynamics of cold trapped ions with application to quantum computation},\ }\href {https://doi.org/10.1007/s003400050373} {\bibfield  {journal} {\bibinfo  {journal} {Applied Physics B}\ }\textbf {\bibinfo {volume} {66}},\ \bibinfo {pages} {181} (\bibinfo {year} {1998})}\BibitemShut {NoStop}%
\bibitem [{\citenamefont {Brown}\ and\ \citenamefont {Carrington}(2003)}]{brown_CUP_2003}%
  \BibitemOpen
  \bibfield  {author} {\bibinfo {author} {\bibfnamefont {J.~M.}\ \bibnamefont {Brown}}\ and\ \bibinfo {author} {\bibfnamefont {A.}~\bibnamefont {Carrington}},\ }\href@noop {} {\emph {\bibinfo {title} {Rotational spectroscopy of diatomic molecules}}}\ (\bibinfo  {publisher} {Cambridge University Press},\ \bibinfo {year} {2003})\BibitemShut {NoStop}%
\bibitem [{\citenamefont {Chalony}\ \emph {et~al.}(2013)\citenamefont {Chalony}, \citenamefont {Barr\'e}, \citenamefont {Marcos}, \citenamefont {Olivetti},\ and\ \citenamefont {Wilkowski}}]{chalony_PRA_2013}%
  \BibitemOpen
  \bibfield  {author} {\bibinfo {author} {\bibfnamefont {M.}~\bibnamefont {Chalony}}, \bibinfo {author} {\bibfnamefont {J.}~\bibnamefont {Barr\'e}}, \bibinfo {author} {\bibfnamefont {B.}~\bibnamefont {Marcos}}, \bibinfo {author} {\bibfnamefont {A.}~\bibnamefont {Olivetti}},\ and\ \bibinfo {author} {\bibfnamefont {D.}~\bibnamefont {Wilkowski}},\ }\bibfield  {title} {\bibinfo {title} {Long-range one-dimensional gravitational-like interaction in a neutral atomic cold gas},\ }\href {https://doi.org/10.1103/PhysRevA.87.013401} {\bibfield  {journal} {\bibinfo  {journal} {Phys. Rev. A}\ }\textbf {\bibinfo {volume} {87}},\ \bibinfo {pages} {013401} (\bibinfo {year} {2013})}\BibitemShut {NoStop}%
\bibitem [{\citenamefont {Zhang}\ \emph {et~al.}(2017)\citenamefont {Zhang}, \citenamefont {Pagano}, \citenamefont {Hess}, \citenamefont {Kyprianidis}, \citenamefont {Becker}, \citenamefont {Kaplan}, \citenamefont {Gorshkov}, \citenamefont {Gong},\ and\ \citenamefont {Monroe}}]{zhang_nature_2017}%
  \BibitemOpen
  \bibfield  {author} {\bibinfo {author} {\bibfnamefont {J.}~\bibnamefont {Zhang}}, \bibinfo {author} {\bibfnamefont {G.}~\bibnamefont {Pagano}}, \bibinfo {author} {\bibfnamefont {P.~W.}\ \bibnamefont {Hess}}, \bibinfo {author} {\bibfnamefont {A.}~\bibnamefont {Kyprianidis}}, \bibinfo {author} {\bibfnamefont {P.}~\bibnamefont {Becker}}, \bibinfo {author} {\bibfnamefont {H.}~\bibnamefont {Kaplan}}, \bibinfo {author} {\bibfnamefont {A.~V.}\ \bibnamefont {Gorshkov}}, \bibinfo {author} {\bibfnamefont {Z.-X.}\ \bibnamefont {Gong}},\ and\ \bibinfo {author} {\bibfnamefont {C.}~\bibnamefont {Monroe}},\ }\bibfield  {title} {\bibinfo {title} {Observation of a many-body dynamical phase transition with a 53-qubit quantum simulator},\ }\href {https://doi.org/10.1038/nature24654} {\bibfield  {journal} {\bibinfo  {journal} {Nature}\ }\textbf {\bibinfo {volume} {551}},\ \bibinfo {pages} {601–604} (\bibinfo {year} {2017})}\BibitemShut {NoStop}%
\bibitem [{\citenamefont {Santra}\ \emph {et~al.}(2022)\citenamefont {Santra}, \citenamefont {Kethepalli}, \citenamefont {Agarwal}, \citenamefont {Dhar}, \citenamefont {Kulkarni},\ and\ \citenamefont {Kundu}}]{Saikat_Riesz_2022}%
  \BibitemOpen
  \bibfield  {author} {\bibinfo {author} {\bibfnamefont {S.}~\bibnamefont {Santra}}, \bibinfo {author} {\bibfnamefont {J.}~\bibnamefont {Kethepalli}}, \bibinfo {author} {\bibfnamefont {S.}~\bibnamefont {Agarwal}}, \bibinfo {author} {\bibfnamefont {A.}~\bibnamefont {Dhar}}, \bibinfo {author} {\bibfnamefont {M.}~\bibnamefont {Kulkarni}},\ and\ \bibinfo {author} {\bibfnamefont {A.}~\bibnamefont {Kundu}},\ }\bibfield  {title} {\bibinfo {title} {Gap statistics for confined particles with power-law interactions},\ }\href {https://doi.org/10.1103/PhysRevLett.128.170603} {\bibfield  {journal} {\bibinfo  {journal} {Phys. Rev. Lett.}\ }\textbf {\bibinfo {volume} {128}},\ \bibinfo {pages} {170603} (\bibinfo {year} {2022})}\BibitemShut {NoStop}%
\bibitem [{\citenamefont {Kethepalli}\ \emph {et~al.}(2024)\citenamefont {Kethepalli}, \citenamefont {Kulkarni}, \citenamefont {Kundu}, \citenamefont {Majumdar}, \citenamefont {Mukamel},\ and\ \citenamefont {Schehr}}]{Jitendra_Riesz_2024}%
  \BibitemOpen
  \bibfield  {author} {\bibinfo {author} {\bibfnamefont {J.}~\bibnamefont {Kethepalli}}, \bibinfo {author} {\bibfnamefont {M.}~\bibnamefont {Kulkarni}}, \bibinfo {author} {\bibfnamefont {A.}~\bibnamefont {Kundu}}, \bibinfo {author} {\bibfnamefont {S.~N.}\ \bibnamefont {Majumdar}}, \bibinfo {author} {\bibfnamefont {D.}~\bibnamefont {Mukamel}},\ and\ \bibinfo {author} {\bibfnamefont {G.}~\bibnamefont {Schehr}},\ }\bibfield  {title} {\bibinfo {title} {Full counting statistics of 1d short range riesz gases in confinement},\ }\href {https://doi.org/10.1088/1742-5468/ad66c5} {\bibfield  {journal} {\bibinfo  {journal} {J. Stat. Mech.}\ }\textbf {\bibinfo {volume} {2024}},\ \bibinfo {pages} {083206} (\bibinfo {year} {2024})}\BibitemShut {NoStop}%
\bibitem [{\citenamefont {Kethepalli}\ \emph {et~al.}(2022)\citenamefont {Kethepalli}, \citenamefont {Kulkarni}, \citenamefont {Kundu}, \citenamefont {Majumdar}, \citenamefont {Mukamel},\ and\ \citenamefont {Schehr}}]{Kethepalli_2022}%
  \BibitemOpen
  \bibfield  {author} {\bibinfo {author} {\bibfnamefont {J.}~\bibnamefont {Kethepalli}}, \bibinfo {author} {\bibfnamefont {M.}~\bibnamefont {Kulkarni}}, \bibinfo {author} {\bibfnamefont {A.}~\bibnamefont {Kundu}}, \bibinfo {author} {\bibfnamefont {S.~N.}\ \bibnamefont {Majumdar}}, \bibinfo {author} {\bibfnamefont {D.}~\bibnamefont {Mukamel}},\ and\ \bibinfo {author} {\bibfnamefont {G.}~\bibnamefont {Schehr}},\ }\bibfield  {title} {\bibinfo {title} {Edge fluctuations and third-order phase transition in harmonically confined long-range systems},\ }\href {https://doi.org/10.1088/1742-5468/ac52b2} {\bibfield  {journal} {\bibinfo  {journal} {J. Stat. Mech}\ }\textbf {\bibinfo {volume} {2022}},\ \bibinfo {pages} {033203} (\bibinfo {year} {2022})}\BibitemShut {NoStop}%
\bibitem [{\citenamefont {Dandekar}\ \emph {et~al.}(2023)\citenamefont {Dandekar}, \citenamefont {Krapivsky},\ and\ \citenamefont {Mallick}}]{Kirone_2023_Riesz}%
  \BibitemOpen
  \bibfield  {author} {\bibinfo {author} {\bibfnamefont {R.}~\bibnamefont {Dandekar}}, \bibinfo {author} {\bibfnamefont {P.~L.}\ \bibnamefont {Krapivsky}},\ and\ \bibinfo {author} {\bibfnamefont {K.}~\bibnamefont {Mallick}},\ }\bibfield  {title} {\bibinfo {title} {Dynamical fluctuations in the riesz gas},\ }\href {https://doi.org/10.1103/PhysRevE.107.044129} {\bibfield  {journal} {\bibinfo  {journal} {Phys. Rev. E}\ }\textbf {\bibinfo {volume} {107}},\ \bibinfo {pages} {044129} (\bibinfo {year} {2023})}\BibitemShut {NoStop}%
\bibitem [{\citenamefont {S.}\ \emph {et~al.}(2021)\citenamefont {S.}, \citenamefont {Huse},\ and\ \citenamefont {Kulkarni}}]{bks21}%
  \BibitemOpen
  \bibfield  {author} {\bibinfo {author} {\bibfnamefont {B.~K.}\ \bibnamefont {S.}}, \bibinfo {author} {\bibfnamefont {D.~A.}\ \bibnamefont {Huse}},\ and\ \bibinfo {author} {\bibfnamefont {M.}~\bibnamefont {Kulkarni}},\ }\bibfield  {title} {\bibinfo {title} {Spatiotemporal spread of perturbations in power-law models at low temperatures: Exact results for classical out-of-time-order correlators},\ }\href {https://doi.org/10.1103/PhysRevE.104.044117} {\bibfield  {journal} {\bibinfo  {journal} {Phys. Rev. E}\ }\textbf {\bibinfo {volume} {104}},\ \bibinfo {pages} {044117} (\bibinfo {year} {2021})}\BibitemShut {NoStop}%
\bibitem [{\citenamefont {Forster}(1975)}]{Forster1975}%
  \BibitemOpen
  \bibfield  {author} {\bibinfo {author} {\bibfnamefont {D.}~\bibnamefont {Forster}},\ }\href@noop {} {\emph {\bibinfo {title} {Hydrodynamic Fluctuations, Broken Symmetry, and Correlation Functions}}}\ (\bibinfo  {publisher} {W. A. Benjamin},\ \bibinfo {address} {Reading, Massachusetts},\ \bibinfo {year} {1975})\BibitemShut {NoStop}%
\bibitem [{\citenamefont {Landau}\ and\ \citenamefont {Lifshitz}(1987)}]{LandauLifshitzFluid}%
  \BibitemOpen
  \bibfield  {author} {\bibinfo {author} {\bibfnamefont {L.~D.}\ \bibnamefont {Landau}}\ and\ \bibinfo {author} {\bibfnamefont {E.~M.}\ \bibnamefont {Lifshitz}},\ }\href@noop {} {\emph {\bibinfo {title} {Fluid Mechanics}}},\ \bibinfo {edition} {2nd}\ ed.,\ \bibinfo {series} {Course of Theoretical Physics}, Vol.~\bibinfo {volume} {6}\ (\bibinfo  {publisher} {Pergamon Press},\ \bibinfo {year} {1987})\BibitemShut {NoStop}%
\bibitem [{\citenamefont {Chakraborti}\ \emph {et~al.}(2021)\citenamefont {Chakraborti}, \citenamefont {Ganapa}, \citenamefont {Krapivsky},\ and\ \citenamefont {Dhar}}]{Subhadip_Blast_2021}%
  \BibitemOpen
  \bibfield  {author} {\bibinfo {author} {\bibfnamefont {S.}~\bibnamefont {Chakraborti}}, \bibinfo {author} {\bibfnamefont {S.}~\bibnamefont {Ganapa}}, \bibinfo {author} {\bibfnamefont {P.~L.}\ \bibnamefont {Krapivsky}},\ and\ \bibinfo {author} {\bibfnamefont {A.}~\bibnamefont {Dhar}},\ }\bibfield  {title} {\bibinfo {title} {Blast in a one-dimensional cold gas: From newtonian dynamics to hydrodynamics},\ }\href {https://doi.org/10.1103/PhysRevLett.126.244503} {\bibfield  {journal} {\bibinfo  {journal} {Phys. Rev. Lett.}\ }\textbf {\bibinfo {volume} {126}},\ \bibinfo {pages} {244503} (\bibinfo {year} {2021})}\BibitemShut {NoStop}%
\bibitem [{\citenamefont {Ganapa}\ \emph {et~al.}(2021)\citenamefont {Ganapa}, \citenamefont {Chakraborti}, \citenamefont {Krapivsky},\ and\ \citenamefont {Dhar}}]{Subhadip_Blast_2021_2}%
  \BibitemOpen
  \bibfield  {author} {\bibinfo {author} {\bibfnamefont {S.}~\bibnamefont {Ganapa}}, \bibinfo {author} {\bibfnamefont {S.}~\bibnamefont {Chakraborti}}, \bibinfo {author} {\bibfnamefont {P.~L.}\ \bibnamefont {Krapivsky}},\ and\ \bibinfo {author} {\bibfnamefont {A.}~\bibnamefont {Dhar}},\ }\bibfield  {title} {\bibinfo {title} {Blast in the one-dimensional cold gas: Comparison of microscopic simulations with hydrodynamic predictions},\ }\href {https://doi.org/10.1063/5.0058152} {\bibfield  {journal} {\bibinfo  {journal} {Physics of Fluids}\ }\textbf {\bibinfo {volume} {33}},\ \bibinfo {pages} {087113} (\bibinfo {year} {2021})}\BibitemShut {NoStop}%
\bibitem [{\citenamefont {Singh}\ \emph {et~al.}(2023)\citenamefont {Singh}, \citenamefont {Chakraborti}, \citenamefont {Dhar},\ and\ \citenamefont {Krapivsky}}]{Singh2023}%
  \BibitemOpen
  \bibfield  {author} {\bibinfo {author} {\bibfnamefont {S.~K.}\ \bibnamefont {Singh}}, \bibinfo {author} {\bibfnamefont {S.}~\bibnamefont {Chakraborti}}, \bibinfo {author} {\bibfnamefont {A.}~\bibnamefont {Dhar}},\ and\ \bibinfo {author} {\bibfnamefont {P.~L.}\ \bibnamefont {Krapivsky}},\ }\bibfield  {title} {\bibinfo {title} {Blast waves in the zero temperature hard sphere gas: Double scaling structure},\ }\href {https://doi.org/10.1007/s10955-023-03127-1} {\bibfield  {journal} {\bibinfo  {journal} {Journal of Statistical Physics}\ }\textbf {\bibinfo {volume} {190}},\ \bibinfo {pages} {118} (\bibinfo {year} {2023})}\BibitemShut {NoStop}%
\bibitem [{\citenamefont {Kumar}\ and\ \citenamefont {Rajesh}(2025{\natexlab{a}})}]{Amit2025_1}%
  \BibitemOpen
  \bibfield  {author} {\bibinfo {author} {\bibfnamefont {A.}~\bibnamefont {Kumar}}\ and\ \bibinfo {author} {\bibfnamefont {R.}~\bibnamefont {Rajesh}},\ }\bibfield  {title} {\bibinfo {title} {Shock propagation following an intense explosion in an inhomogeneous gas: Core scaling and hydrodynamics},\ }\href {https://doi.org/10.1007/s10955-025-03401-4} {\bibfield  {journal} {\bibinfo  {journal} {Journal of Statistical Physics}\ }\textbf {\bibinfo {volume} {192}},\ \bibinfo {pages} {17} (\bibinfo {year} {2025}{\natexlab{a}})}\BibitemShut {NoStop}%
\bibitem [{\citenamefont {Kumar}\ and\ \citenamefont {Rajesh}(2025{\natexlab{b}})}]{Amit2025Splash}%
  \BibitemOpen
  \bibfield  {author} {\bibinfo {author} {\bibfnamefont {A.}~\bibnamefont {Kumar}}\ and\ \bibinfo {author} {\bibfnamefont {R.}~\bibnamefont {Rajesh}},\ }\bibfield  {title} {\bibinfo {title} {Splash in an inhomogeneous gas in one dimension: Exact analysis and molecular dynamics simulations},\ }\href {https://doi.org/10.1007/s10955-025-03502-0} {\bibfield  {journal} {\bibinfo  {journal} {Journal of Statistical Physics}\ }\textbf {\bibinfo {volume} {192}},\ \bibinfo {pages} {119} (\bibinfo {year} {2025}{\natexlab{b}})}\BibitemShut {NoStop}%
\bibitem [{\citenamefont {Cao}\ \emph {et~al.}(2018)\citenamefont {Cao}, \citenamefont {Bulchandani},\ and\ \citenamefont {Moore}}]{Cao2018HardRods}%
  \BibitemOpen
  \bibfield  {author} {\bibinfo {author} {\bibfnamefont {X.}~\bibnamefont {Cao}}, \bibinfo {author} {\bibfnamefont {V.~B.}\ \bibnamefont {Bulchandani}},\ and\ \bibinfo {author} {\bibfnamefont {J.~E.}\ \bibnamefont {Moore}},\ }\bibfield  {title} {\bibinfo {title} {Incomplete thermalization from trap-induced integrability breaking: Lessons from classical hard rods},\ }\href {https://doi.org/10.1103/PhysRevLett.120.164101} {\bibfield  {journal} {\bibinfo  {journal} {Phys. Rev. Lett.}\ }\textbf {\bibinfo {volume} {120}},\ \bibinfo {pages} {164101} (\bibinfo {year} {2018})}\BibitemShut {NoStop}%
\bibitem [{\citenamefont {Bastianello}\ \emph {et~al.}(2020)\citenamefont {Bastianello}, \citenamefont {De~Luca}, \citenamefont {Doyon},\ and\ \citenamefont {De~Nardis}}]{AlvisePRL2020}%
  \BibitemOpen
  \bibfield  {author} {\bibinfo {author} {\bibfnamefont {A.}~\bibnamefont {Bastianello}}, \bibinfo {author} {\bibfnamefont {A.}~\bibnamefont {De~Luca}}, \bibinfo {author} {\bibfnamefont {B.}~\bibnamefont {Doyon}},\ and\ \bibinfo {author} {\bibfnamefont {J.}~\bibnamefont {De~Nardis}},\ }\bibfield  {title} {\bibinfo {title} {Thermalization of a trapped one-dimensional bose gas via diffusion},\ }\href {https://doi.org/10.1103/PhysRevLett.125.240604} {\bibfield  {journal} {\bibinfo  {journal} {Phys. Rev. Lett.}\ }\textbf {\bibinfo {volume} {125}},\ \bibinfo {pages} {240604} (\bibinfo {year} {2020})}\BibitemShut {NoStop}%
\bibitem [{\citenamefont {Bastianello}\ \emph {et~al.}(2021)\citenamefont {Bastianello}, \citenamefont {De~Luca},\ and\ \citenamefont {Vasseur}}]{alvise_review}%
  \BibitemOpen
  \bibfield  {author} {\bibinfo {author} {\bibfnamefont {A.}~\bibnamefont {Bastianello}}, \bibinfo {author} {\bibfnamefont {A.}~\bibnamefont {De~Luca}},\ and\ \bibinfo {author} {\bibfnamefont {R.}~\bibnamefont {Vasseur}},\ }\bibfield  {title} {\bibinfo {title} {Hydrodynamics of weak integrability breaking},\ }\href {https://doi.org/10.1088/1742-5468/ac26b2} {\bibfield  {journal} {\bibinfo  {journal} {Journal of Statistical Mechanics: Theory and Experiment}\ }\textbf {\bibinfo {volume} {2021}},\ \bibinfo {pages} {114003} (\bibinfo {year} {2021})}\BibitemShut {NoStop}%
\bibitem [{sup()}]{supp}%
  \BibitemOpen
  \href@noop {} {}\bibinfo {note} {See Supplemental Material}\BibitemShut {NoStop}%
\bibitem [{\citenamefont {Swope}\ \emph {et~al.}(1982)\citenamefont {Swope}, \citenamefont {Andersen}, \citenamefont {Berens},\ and\ \citenamefont {Wilson}}]{Swope1982}%
  \BibitemOpen
  \bibfield  {author} {\bibinfo {author} {\bibfnamefont {W.~C.}\ \bibnamefont {Swope}}, \bibinfo {author} {\bibfnamefont {H.~C.}\ \bibnamefont {Andersen}}, \bibinfo {author} {\bibfnamefont {P.~H.}\ \bibnamefont {Berens}},\ and\ \bibinfo {author} {\bibfnamefont {K.~R.}\ \bibnamefont {Wilson}},\ }\bibfield  {title} {\bibinfo {title} {A computer simulation method for the calculation of equilibrium constants for the formation of physical clusters of molecules: Application to small water clusters},\ }\href {https://doi.org/10.1063/1.442716} {\bibfield  {journal} {\bibinfo  {journal} {The Journal of Chemical Physics}\ }\textbf {\bibinfo {volume} {76}},\ \bibinfo {pages} {637} (\bibinfo {year} {1982})}\BibitemShut {NoStop}%
\bibitem [{\citenamefont {Kinoshita}\ \emph {et~al.}(2006)\citenamefont {Kinoshita}, \citenamefont {Wenger},\ and\ \citenamefont {Weiss}}]{Kinoshita2006}%
  \BibitemOpen
  \bibfield  {author} {\bibinfo {author} {\bibfnamefont {T.}~\bibnamefont {Kinoshita}}, \bibinfo {author} {\bibfnamefont {T.}~\bibnamefont {Wenger}},\ and\ \bibinfo {author} {\bibfnamefont {D.~S.}\ \bibnamefont {Weiss}},\ }\bibfield  {title} {\bibinfo {title} {A quantum newton's cradle},\ }\href {https://doi.org/10.1038/nature04693} {\bibfield  {journal} {\bibinfo  {journal} {Nature}\ }\textbf {\bibinfo {volume} {440}},\ \bibinfo {pages} {900} (\bibinfo {year} {2006})}\BibitemShut {NoStop}%
\bibitem [{\citenamefont {Pathria}(1986)}]{virial}%
  \BibitemOpen
  \bibfield  {author} {\bibinfo {author} {\bibfnamefont {R.~K.}\ \bibnamefont {Pathria}},\ }\href@noop {} {\emph {\bibinfo {title} {Statistical Mechanics}}},\ International Series in Natural Philosophy\ (\bibinfo  {publisher} {Pergamon Press},\ \bibinfo {year} {1986})\BibitemShut {NoStop}%
\bibitem [{\citenamefont {Castro-Alvaredo}\ \emph {et~al.}(2016)\citenamefont {Castro-Alvaredo}, \citenamefont {Doyon},\ and\ \citenamefont {Yoshimura}}]{Doyon_GHD_PRX}%
  \BibitemOpen
  \bibfield  {author} {\bibinfo {author} {\bibfnamefont {O.~A.}\ \bibnamefont {Castro-Alvaredo}}, \bibinfo {author} {\bibfnamefont {B.}~\bibnamefont {Doyon}},\ and\ \bibinfo {author} {\bibfnamefont {T.}~\bibnamefont {Yoshimura}},\ }\bibfield  {title} {\bibinfo {title} {Emergent hydrodynamics in integrable quantum systems out of equilibrium},\ }\href {https://doi.org/10.1103/PhysRevX.6.041065} {\bibfield  {journal} {\bibinfo  {journal} {Phys. Rev. X}\ }\textbf {\bibinfo {volume} {6}},\ \bibinfo {pages} {041065} (\bibinfo {year} {2016})}\BibitemShut {NoStop}%
\bibitem [{\citenamefont {Bertini}\ \emph {et~al.}(2016)\citenamefont {Bertini}, \citenamefont {Collura}, \citenamefont {De~Nardis},\ and\ \citenamefont {Fagotti}}]{Jacopo_GHD_2016}%
  \BibitemOpen
  \bibfield  {author} {\bibinfo {author} {\bibfnamefont {B.}~\bibnamefont {Bertini}}, \bibinfo {author} {\bibfnamefont {M.}~\bibnamefont {Collura}}, \bibinfo {author} {\bibfnamefont {J.}~\bibnamefont {De~Nardis}},\ and\ \bibinfo {author} {\bibfnamefont {M.}~\bibnamefont {Fagotti}},\ }\bibfield  {title} {\bibinfo {title} {Transport in out-of-equilibrium $xxz$ chains: Exact profiles of charges and currents},\ }\href {https://doi.org/10.1103/PhysRevLett.117.207201} {\bibfield  {journal} {\bibinfo  {journal} {Phys. Rev. Lett.}\ }\textbf {\bibinfo {volume} {117}},\ \bibinfo {pages} {207201} (\bibinfo {year} {2016})}\BibitemShut {NoStop}%
\bibitem [{\citenamefont {Essler}(2023)}]{ESSLER2023GHD}%
  \BibitemOpen
  \bibfield  {author} {\bibinfo {author} {\bibfnamefont {F.~H.}\ \bibnamefont {Essler}},\ }\bibfield  {title} {\bibinfo {title} {A short introduction to generalized hydrodynamics},\ }\href {https://doi.org/10.1016/j.physa.2022.127572} {\bibfield  {journal} {\bibinfo  {journal} {Physica A: Statistical Mechanics and its Applications}\ }\textbf {\bibinfo {volume} {631}},\ \bibinfo {pages} {127572} (\bibinfo {year} {2023})},\ \bibinfo {note} {lecture Notes of the 15th International Summer School of Fundamental Problems in Statistical Physics}\BibitemShut {NoStop}%
\bibitem [{\citenamefont {Doyon}\ \emph {et~al.}(2025)\citenamefont {Doyon}, \citenamefont {Gopalakrishnan}, \citenamefont {M\o{}ller}, \citenamefont {Schmiedmayer},\ and\ \citenamefont {Vasseur}}]{Doyon_GHD_PRX2025}%
  \BibitemOpen
  \bibfield  {author} {\bibinfo {author} {\bibfnamefont {B.}~\bibnamefont {Doyon}}, \bibinfo {author} {\bibfnamefont {S.}~\bibnamefont {Gopalakrishnan}}, \bibinfo {author} {\bibfnamefont {F.}~\bibnamefont {M\o{}ller}}, \bibinfo {author} {\bibfnamefont {J.}~\bibnamefont {Schmiedmayer}},\ and\ \bibinfo {author} {\bibfnamefont {R.}~\bibnamefont {Vasseur}},\ }\bibfield  {title} {\bibinfo {title} {Generalized hydrodynamics: A perspective},\ }\href {https://doi.org/10.1103/PhysRevX.15.010501} {\bibfield  {journal} {\bibinfo  {journal} {Phys. Rev. X}\ }\textbf {\bibinfo {volume} {15}},\ \bibinfo {pages} {010501} (\bibinfo {year} {2025})}\BibitemShut {NoStop}%
\bibitem [{\citenamefont {Santra}\ \emph {et~al.}(2026)\citenamefont {Santra}, \citenamefont {Łebek},\ and\ \citenamefont {Panfil}}]{santra2026crossoverGHD}%
  \BibitemOpen
  \bibfield  {author} {\bibinfo {author} {\bibfnamefont {S.}~\bibnamefont {Santra}}, \bibinfo {author} {\bibfnamefont {M.}~\bibnamefont {Łebek}},\ and\ \bibinfo {author} {\bibfnamefont {M.}~\bibnamefont {Panfil}},\ }\href@noop {} {\bibinfo {title} {Crossover from generalized to conventional hydrodynamics in nearly integrable systems under relaxation time approximation}} (\bibinfo {year} {2026}),\ \Eprint {https://arxiv.org/abs/2603.02158} {arXiv:2603.02158} \BibitemShut {NoStop}%
\bibitem [{\citenamefont {LeVeque}(2007)}]{leveque2007fdm}%
  \BibitemOpen
  \bibfield  {author} {\bibinfo {author} {\bibfnamefont {R.~J.}\ \bibnamefont {LeVeque}},\ }\href@noop {} {\emph {\bibinfo {title} {Finite Difference Methods for Ordinary and Partial Differential Equations: Steady-State and Time-Dependent Problems}}}\ (\bibinfo  {publisher} {Society for Industrial and Applied Mathematics (SIAM)},\ \bibinfo {address} {Philadelphia, PA},\ \bibinfo {year} {2007})\BibitemShut {NoStop}%
\bibitem [{\citenamefont {Press}\ \emph {et~al.}(2007)\citenamefont {Press}, \citenamefont {Teukolsky}, \citenamefont {Vetterling},\ and\ \citenamefont {Flannery}}]{press2007numerical}%
  \BibitemOpen
  \bibfield  {author} {\bibinfo {author} {\bibfnamefont {W.~H.}\ \bibnamefont {Press}}, \bibinfo {author} {\bibfnamefont {S.~A.}\ \bibnamefont {Teukolsky}}, \bibinfo {author} {\bibfnamefont {W.~T.}\ \bibnamefont {Vetterling}},\ and\ \bibinfo {author} {\bibfnamefont {B.~P.}\ \bibnamefont {Flannery}},\ }\href@noop {} {\emph {\bibinfo {title} {Numerical Recipes: The Art of Scientific Computing}}},\ \bibinfo {edition} {3rd}\ ed.\ (\bibinfo  {publisher} {Cambridge University Press},\ \bibinfo {address} {New York},\ \bibinfo {year} {2007})\BibitemShut {NoStop}%
\end{thebibliography}%

\newpage
\setcounter{equation}{0}
\setcounter{figure}{0}

\renewcommand{\theequation}{S\arabic{equation}}
\renewcommand{\thefigure}{S\arabic{figure}}

\onecolumngrid
\clearpage
%\begin{center}
%{\Large{\bf \underline{Supplementary Material} \\Confined Riesz gas: From Newtonian Dynamics to Emergent Hydrodynamics}}
%\end{center}

\begin{center}
    {\bfseries\large
    \underline{Supplemental Material} \\[1ex]
    ``Confined Riesz gas: From Newtonian Dynamics to Emergent Hydrodynamics''}\\[3ex]
    {Indranil Mukherjee,$^1$ Abhishek Dhar,$^1$ and Manas Kulkarni,$^1$}\\[1ex]
    {\small\itshape
    $^1$International Centre for Theoretical Sciences, Tata Institute of Fundamental Research,
    Bangalore 560089, India}
\end{center}

\vspace{0.5cm}
\tableofcontents
\section{Numerical details for the Newtonian dynamics}
\label{sec:newton}
In this section, we describe the numerical methods used for the Newtonian dynamics of the Riesz gas governed by
\begin{align}\label{eq:Riesz_H_supp}
     H({\bf x},{\bf p})=\sum_{i=1}^N \frac{p_i^2}{2m} + \sum_{i=1}^N \frac{m\,\omega^2 x_i^2}{2}  +\frac{J}{2} \sum_{\substack{|i-j|\leq d \\ j\neq i}}^N   \frac{1}{|x_i-x_j|^k}.
\end{align}
Here ${\bf x}=\{x_1,x_2,\ldots,x_N\}$ and ${\bf p}=\{p_1,p_2,\ldots,p_N\}$ denote, respectively, the positions and momenta of the full set of particles. We consider repulsive interactions with $J>0$, set the mass of all particles $m=1$ and take the trap frequency $\omega=1.$ The parameter $d$ controls the interaction range, interpolating from nearest-neighbor ($d=1$) to all-to-all ($d=N-1$) coupling \cite{finite_range_Riesz_MK}. In this work, we focus on the finite-range regime $d\sim O(1),$ and  fix $d=1$ and $k=1.5$. The corresponding equations of motion are 
\begin{gather}\label{eq:Riesz_eom_supp}
    \dot x_i=\frac{p_i}{m},~{\rm and} ~ \dot p_i=-m\omega^2x_i+\frac{Jk}{2}\sum_{\substack{|i-j|\leq d \\ j\neq i}}^N\frac{{\rm sgn}(x_i-x_j)}{|x_i-x_j|^{k+1}}.
\end{gather}
\subsection{Initial state preparation}
\label{supp:ic}

To study the dynamics and approach to the steady state, we initialize the positions of all particles $\{x_i(0)\}$, $i=1,2,\ldots,N$,  from prescribed spatial density profiles $\rho(x)=\sum_{i=1}^N\delta(x-x_i)$ (see below). We perform the simulations, with the total energy of the system fixed to a constant value that is taken to scale as $E\sim O(N^{2\alpha_k+1})$. Specifically, we set $E=\mathcal{C}N^{2\alpha_k+1}$, with the constant $\mathcal{C}=0.5$. 
With the initial positions chosen as above, we compute the total potential energy $E_P$ for this configuration, including both the trapping and interaction contributions in the Hamiltonian given in Eq.~\eqref{eq:Riesz_H_supp}, and determine the total kinetic energy $E_K$ by subtracting $E_P$ from the total energy $E$. Next, 
the velocities $\{v_i(0)\}$ are first chosen from a Maxwell distributions at some temperature and then suitably rescaled such that the total kinetic energy equals $E_k$.  This defines our initial ensemble which is a subset of the constant energy surface. 

For a given set of initial conditions $\{x_i(0),p_i(0)\}$, Eq.~\eqref{eq:Riesz_eom_supp} uniquely determine the subsequent time evolution. To investigate the dynamics and assess thermalization we employ a standard molecular dynamics simulation of Eq.~\eqref{eq:Riesz_eom_supp} based on the symplectic velocity-Verlet integration scheme~\cite{Swope1982} with a very small time increment $dt=10^{-6}$ to ensure numerical accuracy.

The corresponding initial temperature $T_{\rm init}$ is determined from the kinetic energy $E_K$ via the equipartition relation~\cite{virial} 
\begin{equation}
\langle p_i^2\rangle = \Big\langle x_i \frac{\partial H}{\partial x_i}\Big\rangle =T,
\label{eq:supp_ep}
\end{equation}
where we have set $k_B=1$. 
In this work, we consider two different choices of initial density profile $\rho(x,0)$: 
\begin{enumerate}
    \item  {\bf Dome-like equilibrium density profile}: this is obtained by first numerically solving Eq.~\eqref{eq:entropy_min_supp} [see Sec.~\ref{supp-est}] at a prescribed  temperature (and then normalized to $N$). The resulting profile is then used as the probability density for independently drawing the initial particle positions. The sampled positions are then sorted to preserve particle ordering. The velocities are drawn from a Maxwell distribution at a \emph{different} temperature, thereby making the initial ensemble a nonequilibrium one.
    \item {\bf Newton-cradle-type profile~\cite{Kinoshita2006}}: this consists of creating two blobs, each of finite width and separated by a finite distance. The initial positions are drawn independently from a two-component Gaussian mixture: $N/2$ particles are sampled from the Gaussian distribution $\mathcal{G}(x,-d/2,\sigma)$ and the remaining $N/2$ from $\mathcal{G}(x,d/2,\sigma)$, where  $d$ is the center-to-center separation between the two blobs, while $\sigma$ characterizes their respective widths. The corresponding initial density profile is thus 
    \begin{equation}
        \rho(x,0)=\frac{N}{2}\Big[\mathcal{G}(x,-d/2,\sigma)+\mathcal{G}(x,d/2,\sigma)\Big],~{\rm with}~~\mathcal{G}(x,\mu,\sigma)=\frac{1}{\sqrt{2\pi\sigma^2}}~{\rm e}^{-(x-\mu)^2/2\sigma^2},
    \end{equation}
    which satisfies $\int dx~\rho(x,0)=N$, and we choose $d=N/4,\sigma=2.5$. The sampled positions are then sorted to preserve particle ordering. The velocities are chosen from a Maxwell distribution at some temperature. 
\end{enumerate}
\subsection{Time evolution and numerical results}
The density, momentum,  and temperature profiles are constructed from the microscopic data, $\{x_i(t),\{v_i(t)\}$, using the definitions 
\begin{gather}
    \rho(x,t)=\Big \langle \sum_{i=1}^N\delta(x-x_i(t))\Big\rangle \label{eq:profile_rho_defn_supp}\\
    p(x,t)=\rho(x,t)v(x,t)= \Big \langle \sum_{i=1}^N v_i(t)~\delta(x-x_i(t))\Big\rangle \label{eq:profile_v_defn_supp}\\
    \rho(x,t)T(x,t)=\Big \langle \sum_{i=1}^N (v_i(t)-v(x,t))^2~\delta(x-x_i(t))\Big\rangle
\label{eq:profile_T_defn_supp},
\end{gather}
where $\langle\cdots\rangle$ denotes the average over an ensemble of initial conditions drawn from the initial ensemble described in the previous section. 

Note that in Eq.~\eqref{eq:profile_T_defn_supp} the particle velocities are measured relative to the local mean velocity $v(x,t)=p(x,t)/\rho(x,t)$ defined through Eq.~\eqref{eq:profile_v_defn_supp}, so that $T(x,t)$ is the local variance of the velocity distribution rather than its true second moment. This subtraction separates the kinetic energy carried by the coherent (bulk) motion of the fluid from that associated with the random thermal motion of the particles about it. Indeed, since $v(x,t)$ is an ensemble-averaged field, expanding the square in Eq.~\eqref{eq:profile_T_defn_supp} and using Eqs.~\eqref{eq:profile_rho_defn_supp} and \eqref{eq:profile_v_defn_supp}, one obtains
\begin{equation}
\Big\langle \sum_{i=1}^N v_i^2(t)\,\delta\big(x-x_i(t)\big)\Big\rangle = \rho(x,t)\,v^2(x,t) + \rho(x,t)\,T(x,t),
\label{eq:rt}
\end{equation}
i.e., the local kinetic energy density (with $m=1$) splits into a flow contribution $\rho v^2/2$ and a thermal contribution $\rho T/2$.
%Moreover, Eq.~\eqref{eq:profile_T_defn_supp} is Galilean invariant, as required of a thermodynamic temperature: a uniform boost $v_i\to v_i+u$ shifts $v(x,t)\to v(x,t)+u$ but leaves $v_i-v(x,t)$, and hence $T(x,t)$, unchanged. 
Also, in the steady state, where $v(x,t)=0$ and $T(x,t)=T$ is uniform, integrating Eq.~\eqref{eq:profile_T_defn_supp} over $x$ recovers the equipartition relation $\langle p_i^2\rangle=T$ of Eq.~\eqref{eq:supp_ep}.

We present the results of the microscopic simulations for the time evolution of the density profile $\rho(x)$ for the dome-like initial condition (black dashed line) in Fig.~1(a) in the main text. In the simulation, the initial velocities are sampled from a Maxwellian distribution and subsequently rescaled to ensure a fixed total energy $E=\mathcal{C} N^{2\alpha_k+1}$ where we recall that $\mathcal{C}=0.5$. The corresponding initial temperature, determined from the kinetic energy via equipartition [Eq.~\eqref{eq:supp_ep}], is  $T_{\rm init}\approx50$. At long times ($t\sim240$), the system approaches a steady state. The steady-state temperature is again obtained using the equipartition relation [Eq.~\eqref{eq:supp_ep}] applied to the Hamiltonian in Eq.~\eqref{eq:Riesz_H_supp} and the positional degrees of freedom $x_i$. This yields a steady-state temperature $T\approx39.$ The corresponding momentum and temperature profiles, $p(x,t)$ and $T(x,t)$, are shown in Fig.~1(b)-(c) in the main text. As expected, the momentum field vanishes in the steady state and the temperature reaches the expected constant value.
%\begin{figure}[h]
%\centering
%\includegraphics[width= 5.25 cm]{Supp_plots/rho_dome_microa_unscaled.eps}
%\includegraphics[width= 5.25 cm]{Supp_plots/v_dome_micro_unscaled.eps}
%\includegraphics[width= 5.25 cm]{Supp_plots/T_dome_microa_unscaled.eps}
%\includegraphics[width= 0.99\textwidth]{Supp_plots/Fig1_micro.pdf}
%\vspace*{-.2 cm}
%\caption{Newtonian dynamics results, following Eq.~\eqref{eq:Riesz_eom_supp}, for the time evolution of density [Eq.~\eqref{eq:profile_rho_defn_supp}], momentum [Eq.~\eqref{eq:profile_v_defn_supp}], and temperature [Eq.~\eqref{eq:profile_T_defn_supp}] fields, starting from the initial conditions corresponding to a Dome-like initial profile for $N=64$. The steady state is reached at $t\sim240.$ The red squares in (a) are the expected steady-state density profiles obtained from Eq.~\eqref{eq:entropy_min_supp} by rescaling the coordinate $x=yN^{\alpha_k}$. At steady state, the expected momentum profile is zero, shown in (b), and the steady state temperature is $T\sim 39$, shown in the red dashed line in (c). The averages are taken over $10^4$ realizations.}\label{Fig:supp_dome_micro}
%\end{figure}

We present analogous results for a Newton’s-cradle–type initial condition (black dashed line) in Fig.~1 (d)-(f) in the main text. 

To test the independence of the steady state from the initial condition, the total energy of the configuration has been kept fixed to be the same of that for the dome-like profile. The initial velocities (sampled from a Maxwellian distribution) are rescaled accordingly. The system relaxes to a steady state at $t\sim500$. It is expected that the system will reach the same steady state as of the dome-shaped profile. We indeed compute the steady-state temperature, obtained from the equipartition relation and find $T\approx39$, consistent to that of the dome-shaped initial condition. 
%\begin{figure}[h]
%\centering
%\includegraphics[width= 5.25 cm]{Supp_plots/rho_NC_microa_unscaled.eps}
%\includegraphics[width= 5.25 cm]{Supp_plots/p_NC_micro_unscaled.eps}
%\includegraphics[width= 5.25 cm]{Supp_plots/T_NC_microa_unscaled.eps}
%\includegraphics[width= 0.99\textwidth]{Supp_plots/Fig1_micro_NC_supp.pdf}
%\vspace*{-.2 cm}
%\caption{Newtonian dynamics results, following Eq.~\eqref{eq:Riesz_eom_supp}, for the time evolution of density [Eq.~\eqref{eq:profile_rho_defn_supp}], momentum [Eq.~\eqref{eq:profile_v_defn_supp}], and temperature [Eq.~\eqref{eq:profile_T_defn_supp}] fields, starting from the initial conditions corresponding to a  Newton-cradle-type initial profile for $N=64$. The steady state is reached at $t\sim500.$ The red squares in (a) are the expected steady-state density profiles obtained from Eq.~\eqref{eq:entropy_min_supp} by rescaling the coordinate $x=yN^{\alpha_k}$. At steady state, the expected momentum profile is zero, shown in (b), and the steady state temperature is $T\sim 39$, shown in the red dashed line in (c). The averages are taken over $10^4$ realizations.}\label{Fig:supp_NC_micro}
%\end{figure}

\subsection{Estimating the steady state density profile}
\label{supp-est}

The steady-state temperature can be determined analytically from the Riesz gas Hamiltonian in Eq. \eqref{eq:Riesz_H_supp}, following the approach discussed in Ref.~\cite{Riesz_avg_density_SA}. In the regime  $\beta\sim O(N^{-2\alpha_k})$ with $\alpha_k = k/(k+2)$, and $\beta$ being the inverse temperature, the potential energy
\begin{align}\label{eq:Eqntime_dynamics_pot_supp}
     V_k ({\bf x})=\sum_{i=1}^N \frac{m\,\omega^2 x_i^2}{2}  +\frac{J}{2} \sum_{\substack{|i-j|\leq d \\ j\neq i}}^N   \frac{1}{|x_i-x_j|^k}.   
    \end{align}
can be expressed as functional of the density profile $\rho_k(y)$ which takes the following form \cite{Riesz_avg_density_SA,finite_range_Riesz_MK, Jitendra_Riesz_2024}
\begin{equation}\label{Riesz_action_Teff_supp}
\begin{aligned}
\mathcal{V}_k[\rho_k(y)] ={}& \frac{1}{2}\int y^2\rho_k(y)\, dy + J \zeta_d(k)\int\rho_k(y)^{k+1}dy + T_{\rm eff}\int dy \,\rho_k(y)\log\rho_k(y)  - \mu \left(\int \rho_k(y)\, dy -1\right).
\end{aligned}
\end{equation}
In Eq.~\eqref{Riesz_action_Teff_supp}, we have introduced the rescaled coordinate of the position of the particles by a length scale $N^{\alpha_k}$ as $y=x/N^{\alpha_k}$, under which the density transforms as $\rho_k(y)=N^{\alpha_k}\rho_N(x)$, where $\rho_N(x)=\frac{1}{N}\sum_{i=1}^N\delta(x-x_i)$. The normalization conditions $\int_{-\infty}^{\infty} \rho_N(x)dx=1$, and $\int_{-\infty}^{\infty}  \rho_k(y)dy =1$ are thus preserved. The effective temperature in Eq.~\eqref{Riesz_action_Teff_supp} $T_{\rm eff}=T/N^{2\alpha_k} \sim O(1)$. Minimizing the action we get the saddle-point equation 
\begin{align}\label{eq:entropy_min_supp}
    \frac{y^2}{2}+J\zeta_d(k)(k+1)[\rho_k(y)]^k+T_{\rm{eff}} \log [\rho_k(y)]=\mu\,.
\end{align}
The steady-state density profile can be obtained by solving Eq.~\eqref{eq:entropy_min_supp}.  Note that at sufficiently low temperature $\beta >>N^{-2\alpha_k}$, we can neglect the entropy in the action, or equivalently the last term on the l.h.s. of Eq.~\eqref{eq:entropy_min_supp}, and can obtain the exact density profile in this low-temperature limit as mentioned in Ref.~\cite{finite_range_Riesz_MK}. For both initial conditions, we compare the steady-state density profile obtained from the numerical simulation of the Newtonian dynamics with the expected steady state obtained from Eq. \eqref{eq:entropy_min_supp}. The comparison is performed in terms of the unscaled coordinate $x=yN^{\alpha_k}$, and the results are presented in Fig.~1 in the main text.

\section{Derivation of the hydrodynamic Equations}
\label{sec:hydro}
In this section, we provide details of the hydrodynamic equations given in Eqs (14) - (16) in the main text  for the Riesz gas described by the Hamiltonian as in Eq. \eqref{eq:Riesz_H_supp}.
In the absence of the external trap $U(x)$, the system conserves (i) the number of particles $N$, (ii) the total momentum $\sum_{i=1}^N p_i=N\langle p\rangle$, and (iii) the total energy $\sum_{i=1}^N e_i=N\langle e \rangle$, where $p_i$ and $e_i$ is the momentum and energy of the $i$-th particle.  One can then  write a coarse-grained description in terms of the corresponding hydrodynamic fields \cite{LandauLifshitzFluid,Subhadip_Blast_2021,Subhadip_Blast_2021_2}. We consider the density field $\rho(x,t)$, and the momentum density field $p(x,t)$, given in Eq.~\eqref{eq:profile_rho_defn_supp} and Eq.~\eqref{eq:profile_v_defn_supp} respectively. The energy density field $E(x,t)$ can be written as
\begin{equation}
E(x,t)=\rho(x,t)e(x,t)= \Big \langle\sum_{i=1}^N e_i\delta(x-x_i(t))\Big\rangle,  
\end{equation}
where the energy associated with the $i$-th particle is given by 
\begin{align}
 e_i=\frac{p_i^2}{2m}+   \frac{m\,\omega^2 x_i^2}{2}  +\frac{J}{2} \sum_{\substack{|i-j|\leq d \\ j\neq i}}^N   \frac{1}{|x_i-x_j|^k}.    \end{align}
%It is expected that these fields
%$(\rho,p,e)$ should contain the local thermodynamic information at any space-time point $x, t$. 
One can write the total energy per unit mass as 
\begin{equation}
e(x,t)=\epsilon(x,t)+\frac{v(x,t)^2}{2}, 
\label{eq:epm}
\end{equation}
where $\epsilon(x,t)$ in Eq.~\eqref{eq:epm}
denotes the internal energy per unit mass of the system, which also includes the contribution from ideal gas thermodynamics which is $\frac{T}{2}$~\cite{Subhadip_Blast_2021,Subhadip_Blast_2021_2}. The presence of the trap destroys momentum conservation but one can still formulate a hydrodynamic description in terms of the conserved fields $\rho,e$ which are coupled to the slowly decaying momentum field, $p$. The momentum equation gets an extra term related to the external force, $-\partial_x U$, and the resulting hydrodynamic equations are
\begin{eqnarray}
\label{eq:Hydro_eqns1_supp}
&\partial_{t}\rho+\partial_x(\rho v)=0\\
&\label{eq:Hydro_eqns2_supp}
\partial_{t}(\rho v)+\partial_x(\rho v^2+P)+\rho \partial_xU(x)=\partial_x(\xi\partial_x v)\\
\label{eq:Hydro_eqns3_supp}
&\partial_{t}(\rho e)+\partial_x(\rho v e+Pv)=\partial_x(\xi v \partial_x v+\kappa\partial_xT),
\end{eqnarray}
where $\xi$ denotes the bulk viscosity and $\kappa$ is the thermal conductivity of the system.
To solve these hydrodynamic equations [Eqs.~\eqref{eq:Hydro_eqns1_supp} - \eqref{eq:Hydro_eqns3_supp}], one requires an explicit expression for the pressure $P(x,t)$ as a function of the density field $\rho(x,t)$ for the finite-range Riesz gas at finite temperature $T.$ To compute the pressure $P$, we employ the standard thermodynamic relation, 
$P=-\partial_LF\mid_{N,T}$, 
where $F$ is the free energy of the system. For the finite-range Riesz gas with $k>0$, the configurational free energy can be expressed as a functional of the density profile $\rho_N(x)=N\rho(x)$, with $\int \rho_N(x) dx =1$,  as \cite{Riesz_avg_density_SA, finite_range_Riesz_MK}
\begin{align}\label{eq:F_rhoN_supp1}
    F[\rho_N(x)]=\int dx \Big(\frac{x^2}{2}N\rho_N(x)+J\zeta_d(k)N^{k+1}\rho_N(x)^{k+1}+NT\rho_N(x)\log[\rho_N(x)]\Big)
\end{align}
where we define $\zeta_d(k)=\sum_{n=1}^d n^{-k}$, which is the truncated Riemann Zeta function. The three contributions on the r.h.s of Eq.~\eqref{eq:F_rhoN_supp1} correspond, respectively, to the harmonic confinement, the interaction energy, and the entropy term. Rewriting $\rho(x)=N\rho_N(x)$, the local free energy density takes the form: 
\begin{align}\label{eq:f_rho}
    f(\rho)=\frac{x^2}{2}\rho(x)+J\zeta_d(k)\rho(x)^{k+1}+T(x,t)\rho(x)\log\left[\frac{\rho(x,t)}{N}\right]\,,
\end{align}
with the normalization $\int_{-\infty}^{\infty} dx \rho(x)=N$. The pressure is then obtained as:
\begin{align}
P(x,t)=-\partial_L(Lf(\rho))=\rho^2\partial_{\rho}\left(\frac{f(\rho)}{\rho}\right)= J\zeta_d(k)k\rho(x,t)^{k+1}+T(x,t)\rho(x,t)\,.
    \label{eq:Pres}
\end{align}
In the noninteracting limit $J=0$, Eq.~\eqref{eq:Pres} reduces to the ideal gas form $P(x,t)=\rho(x,t) T(x,t)$ \cite{Subhadip_Blast_2021,Subhadip_Blast_2021_2}. Substituting the expression of pressure $P(x,t)$ [Eq.~\eqref{eq:Pres}] into Eq. \eqref{eq:Hydro_eqns2_supp} and using Eq. \eqref{eq:Hydro_eqns1_supp}, Eq. \eqref{eq:Hydro_eqns2_supp} can be recast as
\begin{eqnarray}\label{eq:Hydro_eqns_p_trap_1}
     \rho(x,t)\big( \partial_t +  v(x,t)\partial_x \big)v(x,t)+\partial_x\big[J\zeta_d(k)k\rho(x,t)^{k+1}  + \rho(x,t) T(x,t)\big] =-\rho(x,t)~\partial_xU(x)\,,
\end{eqnarray}
where $U(x)=\frac{x^2}{2}$ is the external harmonic trap potential.

We next determine the  local energy per unit mass using $e(x,t)=\epsilon(x,t)+\frac{v^2(x,t)}{2}$ given in Eq.~\eqref{eq:epm}, where we recall that $\epsilon(x,t)$ is the internal energy per unit mass and the other part is the kinetic contribution from the mean drift. The internal energy, $\epsilon(x,t)$, can be written  as a sum of the contributions from the kinetic and potential energy parts: $\epsilon(x,t)=T(x,t)/2+\epsilon_{pe}(x,t)$, where $T(x,t)$ is the local temperature and $\epsilon_{pe}(x,t)$  is the potential energy contribution. The potential energy contribution can be computed from the free energy density in Eq.~\eqref{eq:f_rho}. This follows on using the thermodynamic relation $\rho \epsilon_{pe}=f+T s$, and noting that the entropy density, $s$, is given by   %Let us denote $u(x,t) = \rho(x,t) \epsilon(x,t)$ as the internal energy density and $S(x,t)$ to be the entropy density.  We then have $u=f+T S$ with
$s=-\big(\frac{\partial f}{\partial T} \big)\mid_\rho$. Hence 
\begin{align}
  \epsilon_{pe}=-\frac{T^2}{\rho} \partial_T \left(\frac{f}{T}\right).  
\end{align}
Using Eq.~\eqref{eq:f_rho}, we thus get
%we then have $S=-\rho(x,t)\log\big(\rho(x,t)/N \big)$. This gives us 
\begin{equation}
\epsilon_{pe}= \frac{x^2}{2}+J\zeta_d(k)\rho^k(x,t).
\end{equation}
%which immediately gives
%\begin{equation}
%\epsilon (x,t)= \frac{u}%{\rho} = \frac{x^2}{2}+J\zeta_d(k)~[\rho(x,t)]^{k}.
\label{eq:supp-eps}
%\end{equation}
%}
Adding the contributions from the kinetic energies (mean flow and fluctuations) the energy per unit mass in Eq.~\eqref{eq:epm} is finally given by
\begin{equation}\label{eq:Hydro_eqns_ext}
    e(x,t)= \frac{x^2}{2}+J\zeta_d(k)~\rho^k(x,t)+\frac{T(x,t)}{2}+\frac{v^2(x,t)}{2}. 
\end{equation}
Substituting Eq.~\eqref{eq:Hydro_eqns_ext} into Eq.~\eqref{eq:Hydro_eqns3_supp} and using Eqs.~(\ref{eq:Hydro_eqns1_supp},~\ref{eq:Hydro_eqns2_supp}), we get after some algebra,
\begin{eqnarray}\label{eq:energy_supp_1_step}
    &J\zeta_d(k)k~\rho^k(x,t)[\partial_t\rho(x,t)+v(x,t)\partial_x\rho(x,t)]+\frac{\rho(x,t)}{2}[\partial_tT(x,t)+v(x,t)\partial_xT(x,t)]\cr
    &+P(x,t)\partial_xv(x,t)+v(x,t)\partial_x(\xi\partial_xv(x,t))=\partial_x(\xi v \partial_x v+\kappa\partial_xT).
\end{eqnarray}
Using the expression of $P(x,t)$, given in  Eqs.~\eqref{eq:Pres}, Eq.~\eqref{eq:energy_supp_1_step} yields 
\begin{equation}\label{eq:energy_supp_1}
\frac{\rho^3(x,t)}{2}(\partial_t+v(x,t)\partial_x)\Bigg(\frac{T(x,t)}{\rho^2(x,t)}\Bigg)=\xi  (\partial_x v)^2+\partial_x(\kappa\partial_x T).
\end{equation}
Collecting ~\eqref{eq:Hydro_eqns1_supp}, \eqref{eq:Hydro_eqns_p_trap_1} and \eqref{eq:energy_supp_1}, we finally have, 
\begin{gather}
\partial_t\rho+\partial_x(\rho v)=0\label{eq:Hydro_eqns_density_supp}\\
    \rho( \partial_t +  v\partial_x )v+\partial_x(J\zeta_d(k)k\rho^{k+1}  + \rho T)+\rho\partial_xU =\partial_x(\xi \partial_xv)\label{eq:Hydro_eqns_momentum_supp}\\
     \frac{\rho^3}{2}\Big(\partial_t+v\partial_x\Big)\Bigg(\frac{T}{\rho^2}\Bigg)=\xi  (\partial_x v)^2+\partial_x (\kappa\partial_x T),  \label{eq:Hydro_eqns_energy_supp}
\end{gather}
where we recall that $\xi$ denotes the bulk viscosity and $\kappa$ is the thermal conductivity of the system.

\section{Numerical details of the hydrodynamic Equations}
\label{sec:hydro-num}
We solve the hydrodynamic equations Eqs.~\eqref{eq:Hydro_eqns_density_supp}-\eqref{eq:Hydro_eqns_energy_supp} numerically using a finite-difference scheme on a one-dimensional grid for both the initial conditions (i) a dome-like profile and (ii) a Newton-cradle-type profile. In both cases, the initial conditions are chosen to match those used in the Newtonian dynamics discussed in Sec.~\ref{supp:ic}.  In the simulation, the space is discretized uniformly with a discretization $dx=0.005$. Time evolution is performed using a fourth-order Runge-Kutta (RK4) method with a time step $dt=10^{-6}$, following standard numerical procedures \cite{leveque2007fdm, press2007numerical}. 
The diffusion constants in the dissipative terms are set to the values $\xi=0.25$ and $\kappa=0.15.$ Numerical convergence is verified by varying $dx$ and $dt$.
Figure 2 (a)-(c) in the main text shows the results of the hydrodynamic simulations for the time evolution of the density  $\rho(x,t)$, momentum $p(x,t)$, and temperature $T(x,t)$ for the dome-like initial profile. The initial conditions are chosen to match those used in the Newtonian dynamics: we consider the same initial distribution $\rho(x)$ with $\int_{-\infty}^{\infty}\rho(x) dx=N$, $p(x,0)=0$, and $T(x,0)=50$, the same initial temperature, determined from the kinetic energy via the equipartition theorem, as discussed in Sec.~\ref{sec:newton}. The density profile from the hydrodynamic simulation relaxes to a steady state at $t\sim 240$. The corresponding hydrodynamic evolution for a Newton’s-cradle–type initial condition is shown in Fig.~2 (d)-(f) in the main text, where the same initial fields $\rho(x,0),p(x,0),T(x,0)$ as in the Newtonian dynamics, are employed. In this case, the system approaches a steady state at $t\sim 500$. For both initial conditions, the steady-state density profiles obtained from the hydrodynamic simulations are compared with the analytical prediction from Eq.~\eqref{eq:entropy_min_supp}. The comparison is performed using the rescaled coordinate $x=yN^{\alpha_k}$. To obtain the steady-state density profile from Eq.~\eqref{eq:entropy_min_supp}, we require the value of the steady-state temperature $T$, obtained using Eq.~\eqref{eq:Hydro_eqns_ext}. Since at steady state, $v(x,t)=0$ and $T(x,t)$ be constant, the total energy can be written, following Eq.~\eqref{eq:Hydro_eqns_ext}, as
\begin{equation}\label{eq:Hydro_T_compute}
    E=\int dx~\frac{x^2}{2}\rho(x)+J\zeta_d(k)\int dx~[\rho(x)]^{k+1}+\frac{T}{2}\int dx~\rho(x),
\end{equation}
where we have used the fact that at steady state $\rho(x,t)\equiv\rho(x)$.
Since in our case the total energy $E$ is conserved between the initial and final states, the steady-state temperature can be determined directly from  Eq.~\eqref{eq:Hydro_T_compute}, which yields $T\approx39$ for both initial conditions. Substituting this value of the steady-state temperature into Eq.~\eqref{eq:entropy_min_supp} then gives the steady-state density profile in the rescaled coordinate. The results are presented in Fig.~2 in the main text for the dome-like and the Newton-cradle-type initial conditions, respectively.

\begin{figure}[t]
\centering
\includegraphics[width= 0.99\textwidth]{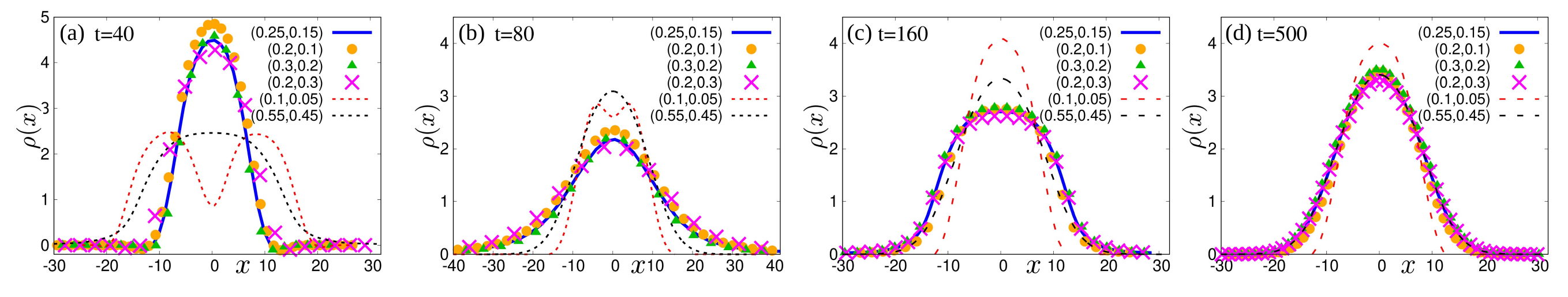}
\includegraphics[width= 0.99\textwidth]{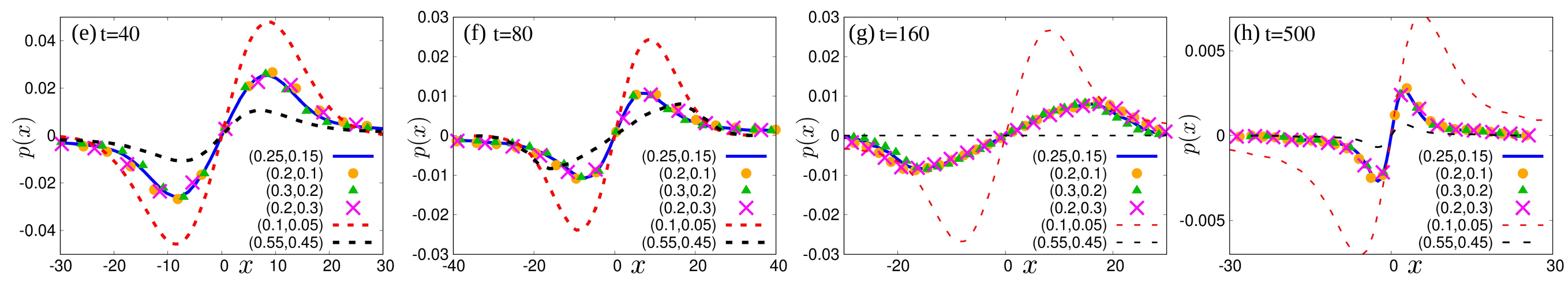}
\includegraphics[width= 0.99\textwidth]{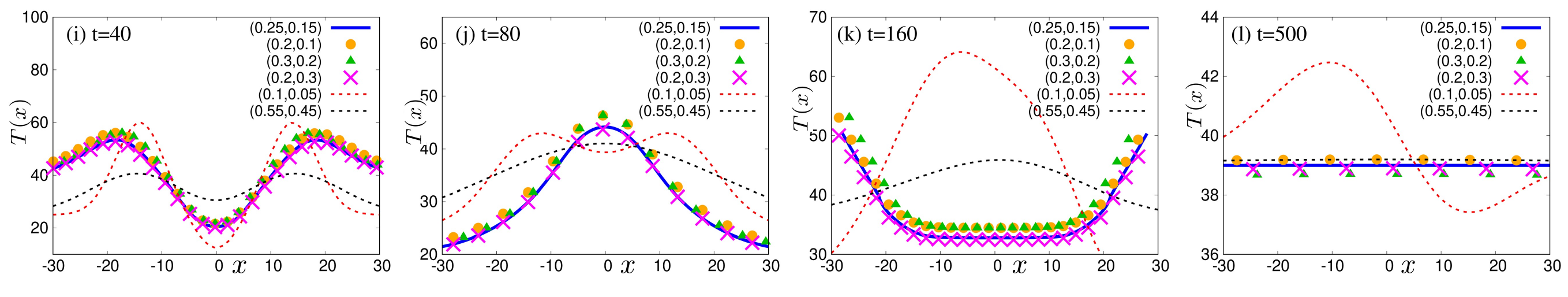}
\vspace*{-.2 cm}
\caption{Comparison of the density $\rho(x,t)$ (top panel), momentum $p(x,t)$ (middle panel), and temperature $T(x,t)$ profiles obtained from hydrodynamic simulations for a  Newton-cradle-type initial condition at times: $t=40, 80,160,500$. Results are shown for different dissipative parameters  $(\xi,\kappa)=(0.2,0.1),~(0.3,0.2),~(0.2,0.3),~(0.1,0.05),(0.55,0.45)$, demonstrating the robustness of hydrodynamic evolution with respect to variations in $\xi, \kappa$. Note that, for $(\xi,\kappa)=(0.1,0.05),(0.55,0.45)$, which are farther from the reference value, the relaxation timescales modify due to the strength of the dissipation, and we do not see the agreement. Recall from Fig.~3 in the main text that the profiles for $\xi=0.25,\kappa=0.15$ (blue) are in agreement with the corresponding profiles from Newtonian dynamics. Therefore, this agreement is preserved across these different parameter sets used.
}\label{Fig:supp_NC_robust}
\end{figure}

\section{Comparison between Newtonian dynamics and hydrodynamics}
\subsection{Quantitative comparison}\label{sec:comp}
In this section we quantitatively compare the results between the Newtonian dynamics [Eq.~\eqref{eq:Riesz_eom_supp}] and  hydrodynamics [Eqs.~\eqref{eq:Hydro_eqns_density_supp}-\eqref{eq:Hydro_eqns_energy_supp}]. For both approaches, we employ identical initial conditions for the dome-shaped and Newton’s-cradle-type profiles, along with a representative choice of dissipative parameters $\xi=0.25,\kappa=0.15$. Figure~3 (a)-(c) in the main text shows this comparison for the density $\rho(x,t)$, momentum $p(x,t)$, and temperature $T(x,t)$ for the dome-like initial condition, over a broad time window spanning from early-time dynamics to the steady state. We observe good agreement between the Newtonian and hydrodynamic results for this case.

For the Newton’s-cradle–type initial condition, we observe a clear interplay between the dissipative transport coefficients and the emergence of hydrodynamic behavior. For very small values of $(\xi,\kappa)$, the 
profiles (density, momentum, and temperature) obtained from
hydrodynamic and Newtonian dynamics  agree at short times but deviate at intermediate times. In this regime, the weak dissipation also slows down relaxation, resulting in a delayed approach to the steady state compared to the microscopic dynamics. In contrast, for moderately large values of ($\xi,\kappa$), the hydrodynamic description fails to capture the very early-time dynamics due to excessive damping. However, beyond an early time (in our case $t \gtrsim 40$ ),  the system crosses over to a regime where the hydrodynamic profiles agree quantitatively with the Newtonian results, and this agreement persists up to the steady state. This behavior is consistently observed across the density $\rho(x,t)$, momentum $p(x,t)$, and temperature $T(x,t)$ fields. 

Figure~3 (d)-(f) in the main text shows this comparison for  Newton’s-cradle–type case over a broad time window, using the same dissipative parameters $(\xi,\kappa)=(0.25,0.15)$ as in the dome-like case.

\subsection{Robustness of hydrodynamic profiles under variations of $\xi$ and $\kappa$}
To assess the robustness of the hydrodynamic description with respect to the dissipative transport coefficients, we solve Eqs.~\eqref{eq:Hydro_eqns_density_supp}–\eqref{eq:Hydro_eqns_energy_supp} for additional parameter sets $(\xi,\kappa)=(0.2,0.1),~(0.3,0.2),~(0.2,0.3)$, close to the reference choice $(0.25,0.15)$ as well as for the values $(\xi,\kappa)=(0.1,0.05),~(0.55,0.45)$ far from the reference. Large variations in $(\xi,\kappa)$ modify the strength of dissipation and can affect the relaxation timescales. Figure~\ref{Fig:supp_NC_robust} shows a comparison of the density $\rho(x,t)$, momentum $p(x,t)$, and temperature $T(x,t)$ profiles at times $t=40,80,160$ and $500$ for a Newton’s-cradle–type initial condition. Across this range of parameters $(\xi,\kappa)=(0.2,0.1),~(0.3,0.2),~(0.2,0.3)$, the profiles remain in close agreement with each other and with those obtained for the reference $(\xi,\kappa)=(0.25,0.15)$, indicating that the hydrodynamic evolution is robust within this reasonably wide parameter window of $(\xi,\kappa)$. Consequently, the agreement with the corresponding Newtonian dynamics is also maintained. In contrast, for parameter choices far from the reference such as 
$(\xi,\kappa)=(0.1,0.05),~(0.55,0.45)$ the relaxation timescales are modified due to weaker and stronger dissipation, respectively. As a result, the corresponding profiles deviate from those obtained with the reference parameters $(0.25,0.15)$.

%For the parameters $(\xi,\kappa)=(0.1,0.05),~(0.55,0.45)$ slightly away from the reference, we see due to relatively small and large dissipation the timescale for the hydrodynamic simulation increases and decreases respectively and the profiles for these parameters are not in agreement with the corresponding profiles for $(\xi,\kappa)=(0.25,0.15)$

For the parameters $(\xi,\kappa)=(0.1,0.05)$, which is much smaller than the reference parameter, we see due to relatively small dissipation the timescale for the hydrodynamic simulation increases and the profiles for these parameters (red dashed in Fig.~\ref{Fig:supp_NC_robust}) are not in agreement with the corresponding profiles for the reference $(\xi,\kappa)=(0.25,0.15)$. On the other hand, for the parameters $(\xi,\kappa)=(0.55,0.45)$, due to relatively large dissipation the timescale for the hydrodynamic simulation decreases (black dashed in Fig.~\ref{Fig:supp_NC_robust}). 

\end{document}